\documentclass[final,12pt]{elsarticle}

\usepackage{amssymb}

\journal{Journal of Non-Crystalline Solids}

\begin{document}

\begin{frontmatter}



\title{Glassy materials for Silicon-based solar panels: present and future}


\author[inst1]{Marcos Paulo Belançon}
\author[inst1]{Marcelo Sandrini}

\affiliation[inst1]{organization={Universidade Tecnológica Federal do Paraná},
            addressline={Campus Pato Branco}, 
            city={Pato Branco},
            postcode={85503-040}, 
            state={Paraná},
            country={Brazil}}

\author[inst2]{Vitor Santaella Zanuto}

\affiliation[inst2]{organization={Universidade Estadual de Maringá},
            addressline={Campus Maringá, Departamento de Física}, 
            city={Maringá},
            state={Paraná},
            country={Brazil}}

\author[inst3]{Robson Ferrari Muniz}

\affiliation[inst3]{organization={Universidade Estadual de Maringá},
            addressline={Campus Goioerê, Departamento de Ciências}, 
            city={Goioerê},
            state={Paraná},
            country={Brazil}}

\begin{abstract}
Glass provides mechanical, chemical, and UV protection to solar panels, enabling these devices to withstand weathering for decades. The increasing demand for solar electricity and the need to reduce anthropogenic carbon emissions demands new materials and processes to make solar even more sustainable. Here, we review the current research to create environmentally friendly glasses and to add new features to the cover glass used in silicon solar panels, such as anti-reflection, self-cleaning, and spectral conversion properties. While several studies have proposed spectral converters and reported information regarding their light-conversion efficiency, there is still a need for a standardized protocol to investigate and compare the impact of these modified materials on the electrical output of photovoltaic systems. In light of these issues, we propose a framework for quantifying parameters that can serve as benchmarks for comparing different cover glasses, which is especially important in the search for a viable spectral converter.
\end{abstract}



\begin{keyword}
Glasses \sep Cover glass \sep Spectral converter \sep Energy
\PACS 0000 \sep 1111
\MSC 0000 \sep 1111
\end{keyword}

\end{frontmatter}



\tableofcontents

\section{Introduction}
\label{intro}
The annual glass consumption worldwide surpassed 21 kg per person in 2014~\cite{Westbroek2021}. Besides traditional applications such as packaging or flat glass for cars and buildings, the glass demand for cover glasses (CG) in solar panels is significant. Silicon-based photovoltaic panels (PV) are already responsible for about 3\% of electricity produced annually worldwide, and this share is expected to grow significantly in the following decades~\cite{Haegel2017,Verlinden2020}. A standard PV produce an electrical output of $\sim$210 $W_p/m^{2}$ from 1000 $W/m^{2}$ of sunlight, which corresponds to efficiencies of about 21\% at the industry level~\cite{ITRPV2022}. As the world transitions to more sustainable energy sources, new PVs are installed as fast as 183 \textit{GW}$_p$ per year,  corresponding to an additional area of about 1 billion square meters. CG demand is high, and the share of bifacial PVs (which may have glass on both the front and back sides~\cite{sinha2021}) is growing and pushing the consumption of float glass by this sector even further.

However, several aspects of the PV technology need further improvements to guarantee its sustainability in the future~\cite{Verlinden2020,Haegel2019,Davidsson2016,Tao2011,Davidsson2017,Bazilian2018}, and some of them are related to the glass ecological footprint~\cite{Burrows2015,FurszyferDelRio2022}, as well as its features, such as UV-filtering, anti-reflective and self-cleaning properties~\cite{Allsopp2020}. Glass makes 67\%-76\% of the total solar panel weight. There is a growing concern about the industrial impact of glass production, which includes significant energy inputs and emissions of about 60 million tons of CO$_2$ equivalent per year~\cite{FurszyferDelRio2022}. From another hand, silicon's characteristic spectral sensitivity limits the efficiency of sunlight to power conversion, and the industry is already reaching the practical limits imposed by the Shockley-Queisser theory~\cite{Shockley1961,Guillemoles2019}. 

In this context, glass science may address these problems and help expand and develop more sustainable technologies, materials, and processes. Here, we review some of the glass research related to this subject, highlighting advances and the most critical challenges associated with PV expansion. We critically discuss these new glassy materials and technologies in light of the terawatt scale of photovoltaic systems worldwide, highlighting aspects that may constrain their adoption, such as inadequate mechanical or chemical properties, abundance of minerals, and cost.

To effectively synthesize and evaluate the vast array of information available, we implemented a systematic approach in our literature review by categorizing the relevant topics into distinct sections. Our objective was to address specific issues of interest to the glass science community, and to establish a cohesive framework for analysing these topics. These topics are organized in terms of the ``energy return on investment'' (EROI) concept~\cite{Hall2014,Bhandari2015,bhandary2015,diesendorf2020,Zhou2018}. The EROI for a solar panel is the sum of energy invested in all materials and processes needed to build the devices, divided by all the energy produced during the panel lifespan. In other words, there are advances that researchers may pursue that will contribute to one or other part of this equation, and some of the most interesting ones are presented and discussed in this work.

In the 1950s, the Pilkington process, or float process, revolutionized the glass-making industry~\cite{Pilkington1969,nascimento2014}. Glass production could grow while enabling glass to be produced faster, with much higher quality and a reduced cost, and these features were necessary to make possible such massive production of PV in the world today. However, even though the knowledge of several families of glasses is abundant, only soda-lime silicate (SLS) glass is adequate to Pilkington's process, and cheap enough to meet the industry needs.

It is important to remember that SLS glass making is an energy-intensive process due to this material high melting temperature ($\sim1500^oC$), which requires about 7-8 GJ/t to be produced~\cite{Westbroek2021}. This value is significantly higher for other materials, such as aluminum (90-100 GJ/t). However, explicitly speaking of the CG, the challenge remains not precisely in how much energy is consumed per ton of glass, but in the amount of material needed in each panel, the growing demand~\cite{ITRPV2022,Burrows2015}, and the difficult task of recycling CG. As this material should have a specific composition, which includes a low iron concentration, it is challenging to perform the recycling process without introducing impurities, namely, iron. In this way, while SLS glass can be indefinitely recycled, the amount of glass mixed with the melt in the float glass production is limited to about $\sim11\%$~\cite{Westbroek2021}, to avoid negative impacts to the final glass sheet.

Considering the vast number of different glasses and treatments applied to their surfaces under investigation by researchers worldwide, here we review some trends that may potentially enhance the EROI of PVs. Several possibilities exist that could either reduce the energy input or increase the energy output of the panel. We will now focus our discussion on those that pertain to the former alternative.

\section{Reducing energy inputs}
At the industry level, glass has become a synonym for SLS glass. However, other glasses, such as borosilicates, are fundamental for some applications due to their improved chemical and thermal resistance. It presents a reduced expansion coefficient, vital for several applications~\cite{youngman2021}, but also preventing it from being thermally toughened~\cite{Allsopp2020}. This fact clearly illustrates that, when developing a new glass system, an improved property often has a downside. 

Anyhow, one can see that basic research plays a critical role in creating an environment that enables breakthroughs to occur. In the case of borosilicate, there is still much research going on about its structure~\cite{manara2009,angeli2012,tostanoski2022}, corrosion~\cite{perea2020}, effects of dopants~\cite{zhu2019} and other basic science studies. In such a context, we are not looking to present a glass ready to replace the SLS in solar panels but to highlight some of the most recent and exciting results in the literature concerning the search for alternatives. 

\subsection{Alternative glass matrix}

Glasses and glass ceramics have been a constant subject of research worldwide. Besides several applications that include lasers~\cite{Guyot2011a}, amplifiers~\cite{belancon2014}, glass fibers~\cite{Dragic2012b,Belancon2013d}, sensors~\cite{Zhong2018,Du2019,Chen2020a} and white-light applications~\cite{Sandrini2018,Chen2014,Wang2020,Kang2008,Silveira2012a,Yi2015,Zhu2017,Zheng2018}, several studies have been developed aiming to apply a glassy material to enhance photovoltaic energy production. In Table \ref{tab:glasses}, we have listed some of these materials recently investigated for this application, as well as their primary raw materials and the respective melting temperature (T$_{Melt}$).

\begin{table}[ht]
    \centering
    \begin{tabular}{|c|c|c|}\hline
         \textbf{Glass} & \textbf{Main Components} & \textbf{T$_{Melt}$} ($^oC$)\\
         \hline
         Aluminosilicates~\cite{sandrini2020,savi2022}&SiO$_2$-CaO-Al$_2$O$_3$&1400-1600\\
         \hline
         YAG Glass-Ceramic~\cite{Tai2018}& SiO$_2$-Al$_2$O$_3$-Y$_2$O$_3$-B$_2$O$_3$ &1500\\
         \hline
         Fluorosilicate~\cite{Fu2012b} & SiO$_2$-Al$_2$O$_3$-CaF$_2$ &1500\\
         \hline
         SLS-Lithium~\cite{Allsopp2020} & SiO$_2$-Na$_2$O-CaO-MgO-Al$_2$O$_3$-Li$_2$O &1450-1480\\
         \hline
         SLS-Titanium~\cite{Bengtsson2022}&SiO$_2$-Na$_2$O-CaO-TiO$_2$&-\\
         \hline
         Alkali alumina-borate GC~\cite{Babkina2019}&Al$_2$O$_3$-B$_2$O$_3$-K$_2$O-Li$_2$O &1400\\
         \hline
         SCS~\cite{Muniz2021,Muniz2021a} & SiO$_2$-Na$_2$O-CaO-Al$_2$O$_3$-CaF$_2$ &1150-1250\\
         \hline
         Recycled SLS~\cite{Gomez2011,Gomez-Salces2016}& SiO$_2$-Na$_2$O-CaO-MgO-Al$_2$O$_3$ & 1100\\
         \hline
         Borates~\cite{YULIANTINI}&B$_2$O$_3$-ZnO-Al$_2$O$_3$-BaO& 1100\\
         \hline
         Phosphates~\cite{Reisfeld1972,REISFELD1978,Reisfeld1980} & NaH$_2$PO$_4$-H$_2$O &1000\\
         \hline
         Lead-Bismuthate~\cite{Pan2000} & Li$_2$O-Bi$_2$O$_3$-PbO &800-1000\\
         \hline
         Lithium-Tellurite~\cite{HumbertodaCunhaAndrade2018} & TeO$_2$-Li$_2$O &850\\
         \hline
         Fluorochlorozirconate~\cite{Leonard2013a}&ZrF$_4$-BaCl$_2$-NaF-AlF$_3$&825\\
         \hline
         Fluorozirconate~\cite{Ahrens2011}&ZrF$_4$-BaF$_2$-NaF-AlF$_3$ &745\\
         \hline
         Tellurite-Tungstate~\cite{belancon2014,Taniguchi2019} & TeO$_2$-WO$_3$-Nb$_2$O$_5$-Na$_2$O &700-800\\
         \hline
         Zinc-Tellurite~\cite{Han2015a,Jiang2016,Garcia2019,Taniguchi2020,Belancon2021} & TeO$_2$-ZnO-Na$_2$O &600-800\\
         \hline
    \end{tabular}
    \caption{Some glassy materials recently investigated for photovoltaic applications and its T$_{Melt}$ value, or range value.}
    \label{tab:glasses}
\end{table}

This list includes materials with chemistry very similar to the commercial SLS glass and some based on entirely different systems. From the point of view of reducing the energy inputs, it would be interesting to develop low-melting temperature materials. However, it seems impossible for large-scale applications to be based on some of those materials in table \ref{tab:glasses}. For example, even though Tellurium is used to produce commercial thin-film technologies such as CdTe solar cells~\cite{Bosio2020}, this mineral is a secondary product in mining~\cite{Graedel2011,Graedel2015a}, and its availability would never allow the production of tellurite glass to be even a fraction of that demanded from PVs.

On the other hand, the upper part of table \ref{tab:glasses} contains a few glass matrices that rely on quite common and abundant minerals.
Yttrium is not as abundant as silicon or aluminum, although it is the second most abundant rare-earth~\cite{Zhang2017c}. Some of these glasses are presented and discussed hereafter, following descending order of melting temperature.

\subsubsection{Aluminosilicates}

The incorporation of aluminum brings substantial modification to silicates in general. Properties such as the elastic module and hardness increase monotonically with alumina content from 30 mol\% to 60 mol\%~\cite{rosales-sosa2016}. Related to this phenomenon, pure silica fibers may have their Brillouin coefficient reduced by two orders of magnitude due to Aluminum~\cite{Dragic2012b}. Aluminosilicate compositions with up to 39\% Al$_2$O$_3$ content~\cite{sandrini2020} have already been used as active media for lasers~\cite{Guyot2011a,DeSousa2003b}, which demonstrates high optical quality as well as efficient luminescent properties that are both required to develop spectral converters (SCs) for photovoltaics. However, the viscosity of the silica melt is significantly reduced by the aluminum~\cite{Hess1996}, and conventional production methods of flat glass are not compatible with these melts~\cite{Dragic2012b}. One exciting possibility is the development of thin films, recently demonstrated by Savi~\cite{savi2022} et al. Though a sophisticated UV-pulsed laser deposition technique was used, it is remarkable that films as thin as 17 nm were obtained, and most properties of the bulk glass samples were preserved. Several complementary works to this study could be exciting, such as investigating the mechanical properties and strength of the film, as well as its production by more straightforward techniques.

\subsubsection{YAG Glass Ceramics}

Yttrium-Aluminum-Garnet (YAG) has been used as active media and phosphor for lighting applications for several decades. Tai et al.~\cite{Tai2018} have prepared glass samples containing Yttrium, which were heat-treated at $\sim750^oC$ to grow YAG nanocrystals inside bulk samples. The resulting Nd$^{3+}$-Yb$^{3+}$ co-doped glass-ceramic demonstrated NIR DC of photons with quantum efficiencies as high as 185\%. One exciting aspect of these materials is the production of nanocrystals at moderately low temperatures, which may favor their production for practical applications. Besides, analyzing this remarkable material on a photovoltaic panel prototype would be essential since the material refractive index and its optical quality may introduce significant reflection losses. As recently demonstrated, such evaluation can be performed under natural sunlight irradiation using an affordable apparatus~\cite{Belancon2021}.

\subsubsection{SLS variations}

Modified SLS glass has also been under investigation aiming at photovoltaic applications. Allsopp et al.~\cite{Allsopp2020} have demonstrated an extensive study of Bi$^{3+}$-Gd$^{3+}$ co-doped SLS glass, which was also slightly modified with the incorporation of Li$_2$O to facilitate the production of flat samples. Miniaturized solar module prototypes fabricated using the most optimal glass samples demonstrated enhanced electricity production, which was explained in terms of the fluorescent dopants incorporated into the glass, even though the authors also pointed out that further measurements are advisable. Remarkably, such samples are similar to the commercial CG, indicating that mass production could be feasible.

Another possibility of improvement of the SLS glass is the mechanical strengthening due to TiO$_2$ incorporation. Bengtsson et al.~\cite{Bengtsson2022} reported a decrease in the alkali diffusion coefficient of SLS glass when Titanium is incorporated by ion exchange. This may be essential to expand the PV lifespan, as Na$^+$ diffusion is one of the leading causes of potential-induced degradation (PID) in these devices. It has been demonstrated that Ti films in SLS glass reduce the PID~\cite{Hara2014}, and the potential to combine this feature with others, such as self-cleaning~\cite{Giolando2016,Pratiwi2020} and anti-reflective~\cite{Khan2020} properties is quite exciting.

\subsubsection{Silicates containing fluorine}

The incorporation of fluorine in silicate glasses has been extensively investigated~\cite{Saito2002,Mukherjee2013,Pei2020}, and it is well-known that this modification reduces both glass transition and peak crystallization temperatures while also improving glass transparency. All these effects can help develop an environmentally friendly CG, as they may reduce energy inputs in glass manufacturing and enhance the sunlight power reaching the solar cells. Muniz, one of the authors of this work, has investigated silicates containing up to $\sim$20\% of CaF$_2$~\cite{Muniz2021}, and rare-earth doped samples~\cite{Muniz2021a,Muniz2021b} based on the system 50SiO$_2$-29Na$_2$O-12.5CaO-7.5CaF$_2$-1Al$_2$O$_3$. Such a high Na concentration, coupled with the fluorine effect, resulted in a significant decrease in the melting temperature to only 1200$^o$C. Down-conversion with up to $\sim$87\% efficiency was achieved in co-doped samples~\cite{Muniz2021a}, and the next steps are underway to evaluate the performance of solar panel prototypes based on this Sodium-Calcium-Silicate (SCS) glass.

Besides the results already demonstrated, this material should have it's chemical and mechanical properties evaluated carefully to check if it can reach the standards needed for applications in solar panels~\cite{Allsopp2020}. Some critical aspects requiring investigation are the chemical stability of the glass. The dissolution of silicates in water~\cite{Kikuchi2022, Guo2022, Gin2021, Geisler2019, Du2019a, Wang2018a} is a complex phenomenon, which is likely to be altered due to the high sodium concentration~\cite{Guiheneuf2017}, and the presence of fluorine, which can make it more challenging to process the material at an industrial scale.

\subsubsection{Other alternatives}

The number of glass systems is limitless and constantly growing and expanding. Many recent works have demonstrated interesting spectroscopic properties~\cite{Seshadri2022, Romero-Romo2021a, Aouaini2022, Kaniyarakkal2023, Bouzidi2022a, Singh2023, Chen2018, Benrejeb2022, Zhang2022, Shi2020}, though, in many of these cases the glass composition depends on a scarce mineral (such as Te), a toxic one (such as Bi, Cd or Pb) or even result in chemically unstable materials. In this regard, though the search for innovative materials for several applications should always be pursued, for PV, more common chemistry, such as those mentioned above, seems to have the best chances. One possibility that does not seem to have been fully explored is the modification of SLS glass, such as proposed by Allsopp et al.~\cite{Allsopp2020}, but also those including surface modification, doping by ion-exchange and others, and we believe researchers worldwide should be encouraged to do so.

\section{Increasing energy outputs}

The conversion of sunlight into electricity is subjected to the Carnot heat engine limit~\cite{Markvart2008,Markvart2016}. As the sun is a black body at 5500$^oC$ and a Si-cell has a working temperature of about 80$^oC$, the Carnot efficiency limit is 94\% for the energy transfer from the sun to the cell~\cite{Chu2012}. However, a Si-cell is not an ideal black body. Due to its spectral sensitivity and internal losses such as electron recombination, the practical efficiency limit for PVs is about 30\%~\cite{Shockley1961,Guillemoles2019,Chu2012,Xu2015}. Considering these constraints, and the relatively complex and expensive processes needed to produce solar cells, it has been fundamental to expand electricity production by maximizing the sunlight reaching the cells.

\subsection{Anti-reflective glass surfaces}

The CG and the encapsulant material in PVs should  be very transparent and exhibit proper refractive indices to reduce reflection losses. Even though sunlight is scattered when reaching the Earth, most of the power produced in PVs is related to the plane-polarized light component. The Fresnel theory states that this component will be reflected in an interface between two mediums, which can be calculated by Equation \ref{eq:fresnel}, 

\begin{equation}
R_s=\left(\frac{n_1 cos \theta_i-n_2\sqrt{1-\left(\frac{n_1}{n_2}sin\theta_i\right)^2}}{n_1 cos\theta_i+n_2\sqrt{1-\left(\frac{n_1}{n_2}sin\theta_i\right)^2}}\right)^2,
\label{eq:fresnel}
\end{equation}
where $R_s$ is the reflection coefficient, $n_1$ and $n_2$ are the refractive indexes of the two mediums, and $\theta_i$ is the incidence angle.

Before reaching the Silicon, sunlight is subjected to air-glass, glass-encapsulant, and encapsulant-silicon interfaces. As the refractive index of Silicon is very high, to avoid $\sim$20\% loss in its interface, an anti-reflective coating on the Silicon surface is mandatory. On the other hand, SLS glass has a refractive index of 1.52, which results in over 4\% loss by reflection for perpendicular incidence of light. Surface texturing of the CG has been explored to produce a refractive index gradient, which further reduces air-glass interface reflection. These structures may enhance the PV's efficiencies by up to 8.7\%~~\cite{Kim2020a}. Some theoretical work has even proposed surface texturing to increase the panel's emissivity~\cite{Zhou2021} as a pathway to reduce its temperature.

AR coatings based on ``Moth-eye'' and multiple interference films have been investigated, and several techniques to produce them have already been demonstrated~\cite{Buskens2016,Lobmann2018}. These and other results are fascinating; however, as PVs should withstand weathering for at least a few decades, it is fundamental to investigate the durability of these coatings under the most distinct climatic and meteorological conditions and specific conditions caused by regionally located situations.~\cite {Morales2018,Womack2019}.

\subsection{Self-cleaning and multifunctional glass surfaces}

PVs are supposed to produce as much electricity as possible during their lifetimes. These devices are installed literally in all continents of the Earth and, in this sense, are subject to a wide range of environments, wind, storms, etc. Soiling is an important issue~\cite{Maghami2016a} as it may significantly reduce the amount of light reaching the solar cells inside the panel. This has fuelled the development of self-cleaning surfaces.

In some cases, rain is well distributed throughout the year, which can be enough to keep PVs satisfactorily clean. However, in some environments, such as dry or icy ones, even though one may have an excellent potential for solar power production, keeping the CG surfaces clean can be challenging as water may not be available~\cite{Syafiq2018,Lu2020}. In the context discussed in this work, some exciting results have proposed multifunctional coatings exhibiting self-cleaning, anti-reflective, and even luminescent properties~\cite{cheng2021,biswas2022}. We will come back to this subject on section \ref{sec:multi}, but next we review some research on spectral converters.

\subsection{Spectral converters (SC)}

In a commercial PV, most of the sunlight is converted into heat~\cite{Phys2022}, and the spectral mismatch between the sunlight and silicon	 sensitivity plays a central role in this inefficiency~\cite{Shockley1961,Guillemoles2019,Chu2012,Xu2015}. The standard 1.5G solar spectrum has an intensity of 1000 $W/m^2$. Photons below the silicon bandgap ($\sim$1100-2200~nm) account for 164 $W/m^2$~\cite{Huang2013}. Theoretically, these photons can be up-converted to higher energy ones, enabling additional sunlight to produce electricity. However, up-conversion efficiency is inherently low~\cite{tawalare2021}, and after extensive research on the subject in the last decades~\cite{Ghazy2021}, there is no clear evidence of a feasible up-converter for PVs. Most experiments claiming to measure an increment in electricity output due to up-converter materials were performed in impractical conditions, often based on laser illumination and/or highly concentrated light
~\cite{Khare2020}. In this way, the stokes shift is far more likely to occur than the anti-stokes shift, hence we will focus here on the spectral conversion of the above bandgap photons only.

Each incoming photon may produce a maximum of one free electron in the Si-cell by creating an electron-hole pair. Above bandgap photons, however, have excess energy that will be wasted. About 149 $W/m^2$ of the incoming sunlight consists of UV-blue photons ($\sim$300-450~nm) with at least twice the silicon bandgap energy~\cite {Huang2013}. Besides this significant loss due to energy excess, these high-energy photons are also linked to solar panel degradation~\cite{Lin2016c, Schlothauer2012, Katayama2019, Zhang2017b,francisnara2022, Gopalakrishna2019}. The electricity output could be increased if these photons were converted or split into two half-energy photons. This could modify the spectrum to enrich the NIR part, where PV is more sensitive, reducing the heat associated with the electron-hole pair creation or even increasing the number of photons reaching the solar cell (if the SC exhibits quantum efficiency higher than one).

Most of the proposed SCs are based on rare-earth active ions, though there is plenty of research on other dopants, such as transition metals or metallic nanoparticles~\cite{dechaoyu2022,satpute2022}. Two main approaches concerning SC exist: developing CGs containing optically active ions and producing optically active films on top of a standard CG. Although the list of dopants being investigated is long and has been reviewed recently~\cite{dechaoyu2022,satpute2022}, here we provide a short description of the most common ones, highlighting some key aspects of the complex task of developing a viable SC.

\subsubsection{Main rare-earth dopants for spectral conversion}

\textit{Ce$^{3+}$}

Cerium is one of the most abundant rare-earths and is widely used in industrial applications such as catalysis, UV-blocking agent in glasses, etc. When incorporated into glass, it can exist in two different valence states, either Ce$^{3+}$ or Ce$^{4+}$~\cite{Taniguchi2020}. As Ce has the atomic number 58, it will have 54 electrons in its oxidized state and present full electron shells and no optical transitions.

The additional electron in the Ce$^{3+}$ ion, though, has a 4f-5d parity allowed transition~\cite{lupei2014}, which is very sensitive to the host crystal field. Absorption bands of Ce$^{3+}$ are often observed in the UV-blue range~\cite{Teng2020a}, while emission bands have been reported ranging from UV-blue~\cite{Annapurna2004,pullaiah2022} to the yellow-red parts of the visible spectrum~\cite{Andrade2009}. In this way, to develop SCs for PVs, Cerium is often used to absorb UV-blue light. In most cases, it is designed to work coupled with some other ion such as Nd$^{3+}$~\cite{taizheng2015,wangqiu2015} or Yb$^{3+}$~\cite{zhoulei2022, Reddappa2021,pathak2017} which are well-known for their emission lines near the silicon sensitivity peak~\cite{Phys2022}.

One of the drawbacks reported with Cerium-doped glasses is the difficulty of controlling its valence states. In most cases, the final material will have a mix of both states~\cite{sontakke2016,ranasinghe2022,kaewnuam2022}. For PV applications, Ce$^{4+}$ may be an excellent UV-blocking agent, though it may prevent spectral conversion~\cite{Taniguchi2020}.\\

\textit{Pr$^{3+}$}

The Pr$^{3+}$ energy diagram is broadly investigated as it has several absorption and emission bands between the blue and the near-infrared, including an emission line near the silicon sensitivity peak around 1$\mu m$~\cite{belancon2014}. In practice, most of the blue light absorbed by this ion will suffer a slight Stokes shift and result in visible emission, which will barely benefit the PV efficiency. Similar to the case of Cerium, Praseodymium has been proposed to be used alongside a NIR emitter, such as Yb$^{3+}$~\cite{song2012}. Such mechanism of spectral conversion has been demonstrated experimentally~\cite{zhaozhao2020}. The energy diagram of this ion results in a relatively narrow absorption band in the blue region due to ground state absorption to the $^3$P levels. Because of that, besides co-doping schemes with a NIR emitter, Pr$^{3+}$ has also been used as co-dopant with Ce$^{3+}$~\cite{zhangcui2016}.

However, as we have pointed in a recent work~\cite{Belancon2021}, one should not neglect the negative impact of any ion	 absorption bands on a solar cell performance. Even if the spectral conversion can be observed, it has no straightforward relation to the overall efficiency of a solar cell prototype. In the case of Praseodymium doped and co-doped samples, electrons excited to the $^3$P levels may result in quantum-cutting or down-conversion. However, the lower energetic $^1$D$_2$ level is expected to reduce the yellow-red ($\sim590$ nm) transmission through the sample. Even though resonant emission from this level is expected ($\sim$612 nm), it will also result in the emission of low-energy photons, including some with energy below the silicon bandgap~\cite{belancon2014} (for example at $\sim$1480 nm).\\

\textit{Eu$^{2+}$}

Europium-doped glasses will more often result in Eu$^{3+}$ rich materials, and this valence state provides a mechanism to obtain intense red emission~\cite{Sandrini2018,Garcia2019,Muniz2021b,Meejitpaisan2021}. This ion can sometimes be found or reduced to Eu$^{2+}$~\cite{Muniz2021a,Luo2019}, which has spectroscopical properties similar to Ce$^{3+}$. Indeed, Dorenbos~\cite{dorenbos2003,dorenbos2003a} have demonstrated a strong correlation between the spectroscopic properties of these two ions. Also similar to the case of Cerium, Europium has been incorporated into several different materials where often a NIR emitter co-dopant is used, such as Pr$^{3+}$~\cite{chenwang2017}, Dy$^{3+}$~\cite{wang2015}, Nd$^{3+}$~\cite{Muniz2021a,talewar2016,luo2019b} and Yb$^{3+}$~\cite{taizheng2015}.\\

\textit{Nd$^{3+}$}

As previously mentioned, Nd$^{3+}$ has been used as a NIR emitter in materials proposed as SCs for PVs~\cite{Muniz2021a,taizheng2015,Meejitpaisan2021,talewar2016,luo2019b,zhoutenk2011}. However, this ion has also been a donor to Yb$^{3+}$ ions~\cite{savi2022,Tai2018}. Though it could be theoretically possible to perform some spectral conversion using Nd$^{3+}$ single-doped samples, in practice, the ion has several absorption lines in the visible and NIR that are likely to have some negative impact in the spectrum reaching the Si-cell. We will be back to this question hereafter.\\

\textit{Yb$^{3+}$}

Yb$^{3+}$ is probably the most investigated NIR emitter for PVs~\cite{savi2022,Tai2018,zhaozhao2020,Romero-Romo2021,Zhou2016,Elleuch2015,Bouzidi2022,Saad2022}. The main reason for that relies on the simple energy levels diagram of this ion, which has absorption bands (between 850-1000 nm) and intense emission (between 970-1050~nm) in the NIR range. Though absorption in this region may not be beneficial for PVs~\cite{Allsopp2020,kempe2009}, Yb$^{3+}$ can be excited by energy transfer from several rare-earths and transition metals~\cite{Saad2022,Dan2023}. In the next section, we discuss the challenges related to choosing dopants.
\\

\textit{Selecting rare earth elements for spectral converters.}
\\

\begin{figure}[htbp]
    \centering
    \includegraphics[scale=0.8]{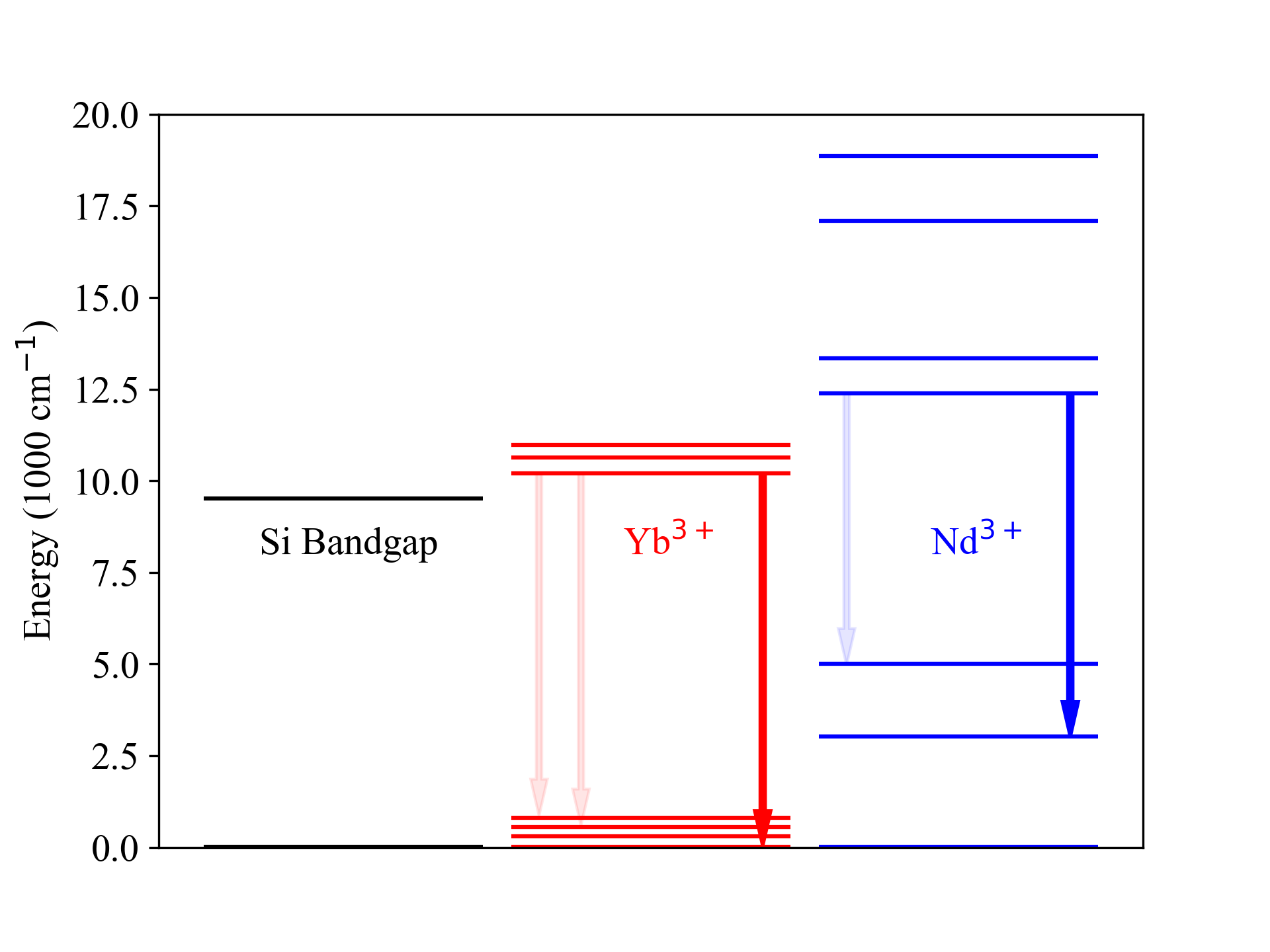}
    \caption{Representative energy levels diagram of NIR emitters for spectral conversion. Light color and darks color arrows indicate emission lines below and above the Si bandgap, respectively.}
    \label{fig:diagram}
\end{figure}

As we have pointed out, several rare-earth doped materials have been investigated as a mechanism to achieve spectral conversion. In figure \ref{fig:diagram}, we show a simplified representation of the main energy levels in Yb$^{3+}$ and Nd$^{3+}$, which play an important role in the development of SCs.

Absorption of photons with wavelengths higher than ~1100 nm is unharmful to the electrical output of PVs, as these photons have significantly low energy. As we have mentioned, the up-conversion from these photons seems unlikely and is not within the scope of this work. On the other hand, absorption in the range of 300-1100 nm is desirable only if we can increase the electrical output of the PV after the spectral modification. As the optimum sensitivity of PVs is found near the Si bandgap, there is no point in having an SC absorbing energy in this region.

Yb$^{3+}$ is transparent in almost all silicon sensitivity range and exhibits a resonant absorption/emission matching the PV sensitivity peak. On the other hand, Nd$^{3+}$ exhibits many absorption peaks along the UV-visible range, presenting a solid emission in the NIR region. However, the same excited level related to the emission at $\sim$1064 nm is also the origin of the emission at $\sim$1350 nm, below the bandgap. Nd$^{3+}$ is also a strong absorber between 730-830 nm, a region where the down-conversion of a photon is not favorable for PV.

To the best of our knowledge, the research on both Yb$^{3+}$ and Nd$^{3+}$ based SCs has yet to consider the negative impacts of this ion on the overall efficiency of PVs. Most of the research has focused on optimizing the dopant concentrations to obtain higher quantum efficiencies in the conversion of photons; however, the PV electrical output is not a function of this single variable, and neither correlates straightforwardly to it. In some materials exhibiting high conversion efficiencies, a high dopant concentration of Yb$^{3+}$ or Nd$^{3+}$ is used, which could reduce the electricity output of Si-cells.

It is essential to highlight that several reports on high conversion efficiencies have been demonstrated in the last few years. However, none have included experimental measurements to quantify a prototype's electrical output under sunlight irradiation. As recently shown in a Pr$^{3+}$-doped tellurite glass~\cite{Belancon2021}, it is possible to detect rare-earth emission bands under natural light, and we believe the SCs should be evaluated under this situation. In section \ref{model}, we propose an analysis considering this goal.

As one can imagine, the list of ions that could be explored to provide UV-Blue absorption and NIR emission is extensive. Other rare-earths such as Tb$^{3+}$~\cite{Mattos2022}, Er$^{3+}$~\cite{Yang2014}, Dy$^{3+}$~\cite{Kadam2021}, Ho$^{3+}$~\cite{Jia2018}, Tm$^{3+}$~\cite{Zhou2016} have also been investigated. However, as we have discussed, a ``too rich'' energy diagram will often result in a loss of sunlight power somewhere in the spectrum by one of two mechanisms: 1) undesirable absorption of photons or 2) undesirable emission of photons without overall gains for the electricity output.

\subsubsection{Non-rare-earth ions for spectral conversion}

Several transition metals (TMs) have also been investigated as active ions for spectral conversion~\cite{Allsopp2018,Fujita2018,Peng2009,Ghosh2015}. Some of the most exciting results reported concern the possibility of replacing cerium by a TM to block UV transmission, as Cerium may react with traces of iron in the SLS glass under UV radiation. The result, in this case, is an increase of the Fe$^{2+}$ concentration in the CG~\cite{kempe2009}, a NIR absorber that will have a deleterious effect on the PV electricity output. It has been demonstrated by using TMs~\cite{Allsopp2018} that significant UV blocking capabilities can be achieved, with the bonus that some TMs exhibit VIS/NIR emission, and down-conversion can potentially be achieved as well.

Silicon nanocrystals are another alternative to develop spectral converters~\cite{Svrcek2004}. A few different techniques to produce silicon nanocrystals are available~\cite{Luxembourg2014,Yuan2020,Das2021}, and some works have even demonstrated solar panel prototypes with enhanced performance~\cite{Luxembourg2014,Das2021}. Authors have explained these promising results due to the conversion of UV photons into visible ones and reduced surface reflectance. Some undesirable aspects reported include a reduced transmittance of the host material with increasing nanocrystal concentration. Silicon is an abundant mineral, and some authors~\cite{Das2021} have demonstrated its deployment using a silica gel as a host, which could be a path to improve the efficiency of solar panels on-field.

\subsubsection{A benchmark framework for spectral converters}
\label{model}
To the best of our knowledge, there is no standardized test to measure the performance of SCs. Indeed, as we have discussed, most works proposing an SC have some information about the conversion efficiency of the absorbed light. However, they lack specific measurements of the effect of the material on the electrical output of a PV. On the other hand, we recognize how difficult it is to produce solar cell prototypes covered by the SC. It can be a resource and time-consuming task that requires a different set of skills and equipment and could introduce uncertainties in comparing different SCs. With this aim, we propose a framework for quantifying some parameters that could be used as a benchmark to compare SCs.

Our approach is based on the following considerations:
\begin{enumerate}
    \item The Global 37$^o$ tilt ASTM-G173-03 sunlight spectrum (1.5G spectrum) is used as the reference for standard sunlight illumination.
    \item The Thorlabs S120VC Si sensor responsivity spectrum is used to represent a Si solar cell spectral sensitivity.
    \item The product of these two spectra represents the solar cell current output per nm under sunlight illumination.
\end{enumerate}

\begin{figure}[htbp]
    \centering
    \includegraphics[scale=0.6]{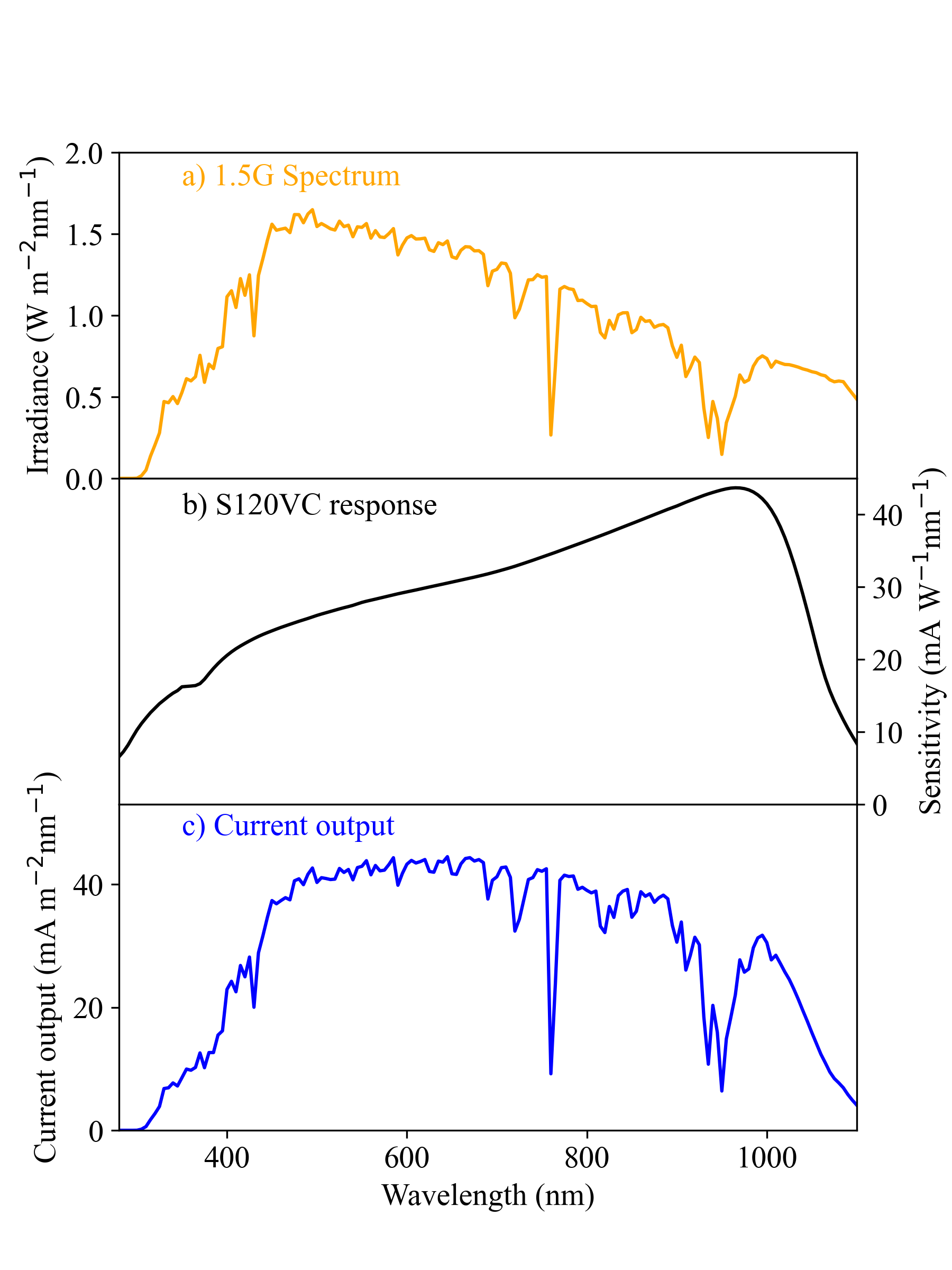}
    \caption{a) 1.5G irradiance spectrum and b) spectral sensitivity of commercial S120VC silicon sensor. The combination of these two results in a curve of c) current output of the sensor per nm of the 1.5G spectrum.}
    \label{fig:scheme}
\end{figure}

These three curves enumerated above are shown in figure \ref{fig:scheme}, for the range 280-1100~nm, with a resolution of 5~nm. The area under these curves was calculated for different sections of the data, and all of them were performed using the trapezoidal rule. The integral of the 1.5G spectrum in this range and resolution corresponds to 805.65~W/m$^2$, and as one can see, the sunlight intensity peaks at $\sim$500~nm. On the other hand, the Si sensitivity peaks at $\sim$1000~nm. At the bottom of the figure, we show the resulting ``current output spectrum'' obtained by multiplying the sunlight irradiance times the sensitivity curve. 

The total area under the last curve corresponds to 24159 mA/m$^2$, and it is essential to highlight the ``flat'' region, between 450-900~nm, which is responsible for about 75\% of the total current output ($\sim$17917 mA/m$^2$). One fundamental insight of this approach is that, in the flat region of the current curve, its value is $\sim$40~$mA~m^{-2}nm^{-1}$. This means a production of about 200~mA/m$^2$ for each 5~nm interval. As one can see, even a narrow absorption band may significantly affect electricity production depending on the SC thickness. This way, if we have a 5~nm narrow band in the 450-900~nm range that absorbs 50\% of the incoming light, this will reduce the current output per square meter in $\sim$100~mA/m$^2$, which corresponds to about 0.41\% of the current output under direct sunlight.

\begin{figure}[htbp]
    \centering
    \includegraphics[scale=0.6]{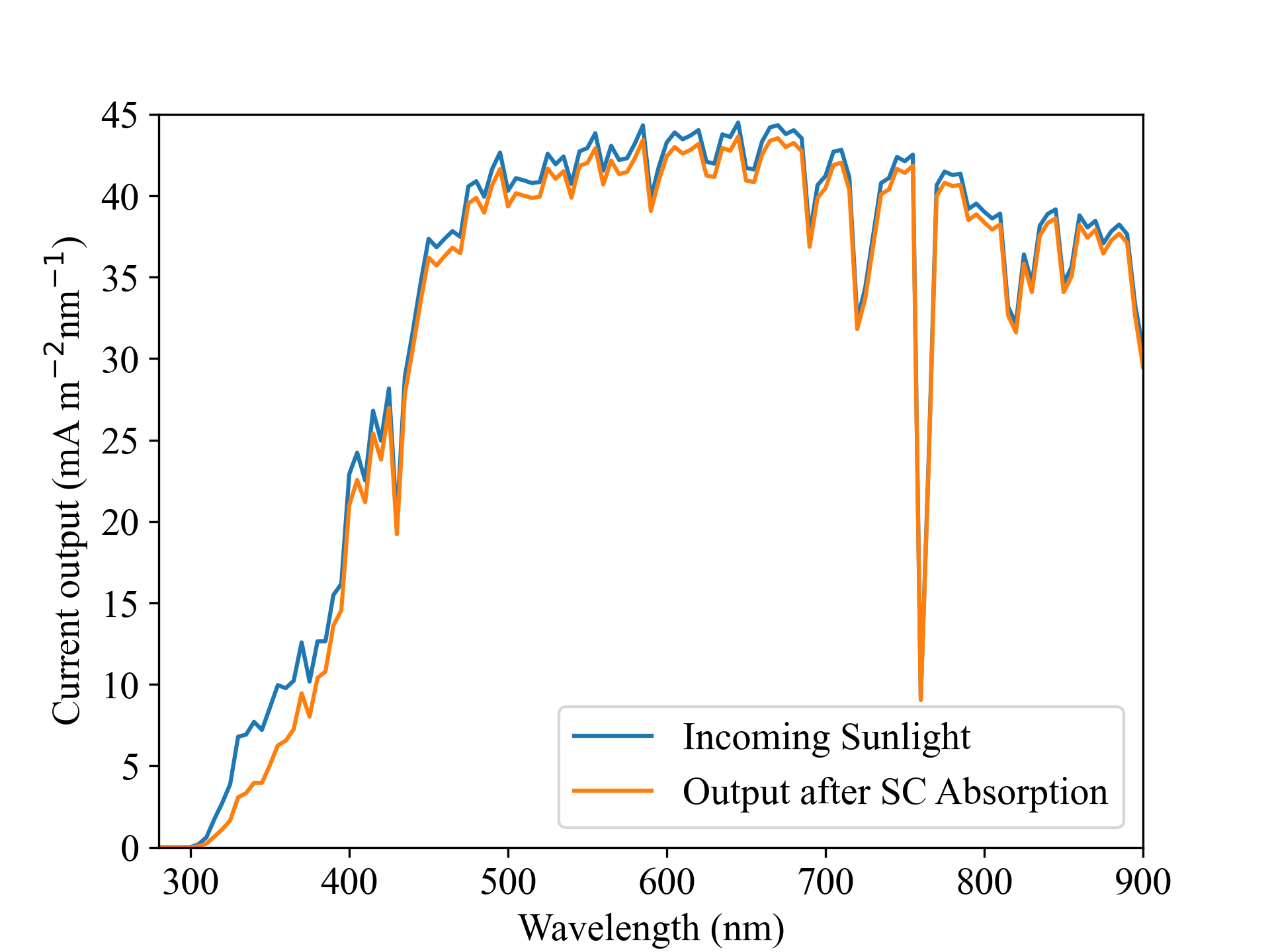}
    \caption{Calculated current output under 1.5G sunlight spectrum and the effect of a 0.1 mm thick Ce-doped xerogel based film~\cite{Sandrini2023}.}
    \label{fig:SolGelSC}
\end{figure}

Using the material absorption spectrum and this current output reference curve, one can estimate the negative impact of an SC on the electricity output. To illustrate that, we evaluated the absorption data in the range 280-900~nm from a 0.1~mm thick Cerium-doped xerogel~\cite{Sandrini2023}, and the results are shown in Figure \ref{fig:SolGelSC}.

Considering the sunlight spectrum in that range, this material applied to a PV would reduce the current output from 20114 to 19471 mA/m$^2$ due to its absorbance. From this 643~mA/m$^2$ loss, 265~mA/m$^2$ are related to wavelengths below 420~nm, demonstrating the reduction, in this case, is more significant in the UV range. This illustrates how to quantify the losses introduced by the material, though, for an SC, it is also necessary to evaluate the potential gains which can be introduced. A viable SC must add enough spectral conversion to overcome its absorption losses. In the next section, we introduce a model for an ``ideal SC'', which could be used as a reference scale for different materials proposed as SCs.

\subsubsection{The ideal SC}

We can think on different types of SCs, made from different compositions and having specific spectroscopic characteristics. However, to develop our reference for an ideal SC, it is enough to consider that it exhibits the following attributes:
\begin{itemize}
    \item Full absorption of light below a certain wavelength;
    \item Full transparency above this same wavelength;
    \item Convert all the absorbed radiation into wavelengths near the Si sensitivity peak;
    \item Do not introduce any other loss due to the sunlight reflection or scattering;
\end{itemize}

The immediate implication is that this ideal SC will have more photons emitted than absorbed because we are converting high-energy photons into NIR ones without generating any phonons. Though a quantum efficiency as high as this is unlikely, the above considerations are enough to provide a simple and helpful reference. On the other hand, the wavelength that delimits the absorption and transparency of the SC is arbitrary. Though, as we discussed before, 75\% of the current output in our framework originates from sunlight in the range 450-900~nm, and in this way, it seems reasonable not to consider absorption near that range. In this context, we chose the wavelength limit between absorption/transparency for our ideal SC to be 420~nm.

The Si sensitivity response we are considering, as shown in Figure \ref{fig:scheme}, peaks around 965~nm. An ideal SC should profit from that by emitting radiation near it. Here again, there is room for discussion on the shape and width of the SC luminescence band. A ``laser-like'' luminescence is quite unlikely, but on the other hand, a too-broad band would mean some photons falling below the Si bandgap. In this way, we choose a simple flat emission band between 940-990~nm, which is 50~nm wide and centered at 965~nm.

\begin{figure}[htbp]
    \centering
    \includegraphics[scale=0.6]{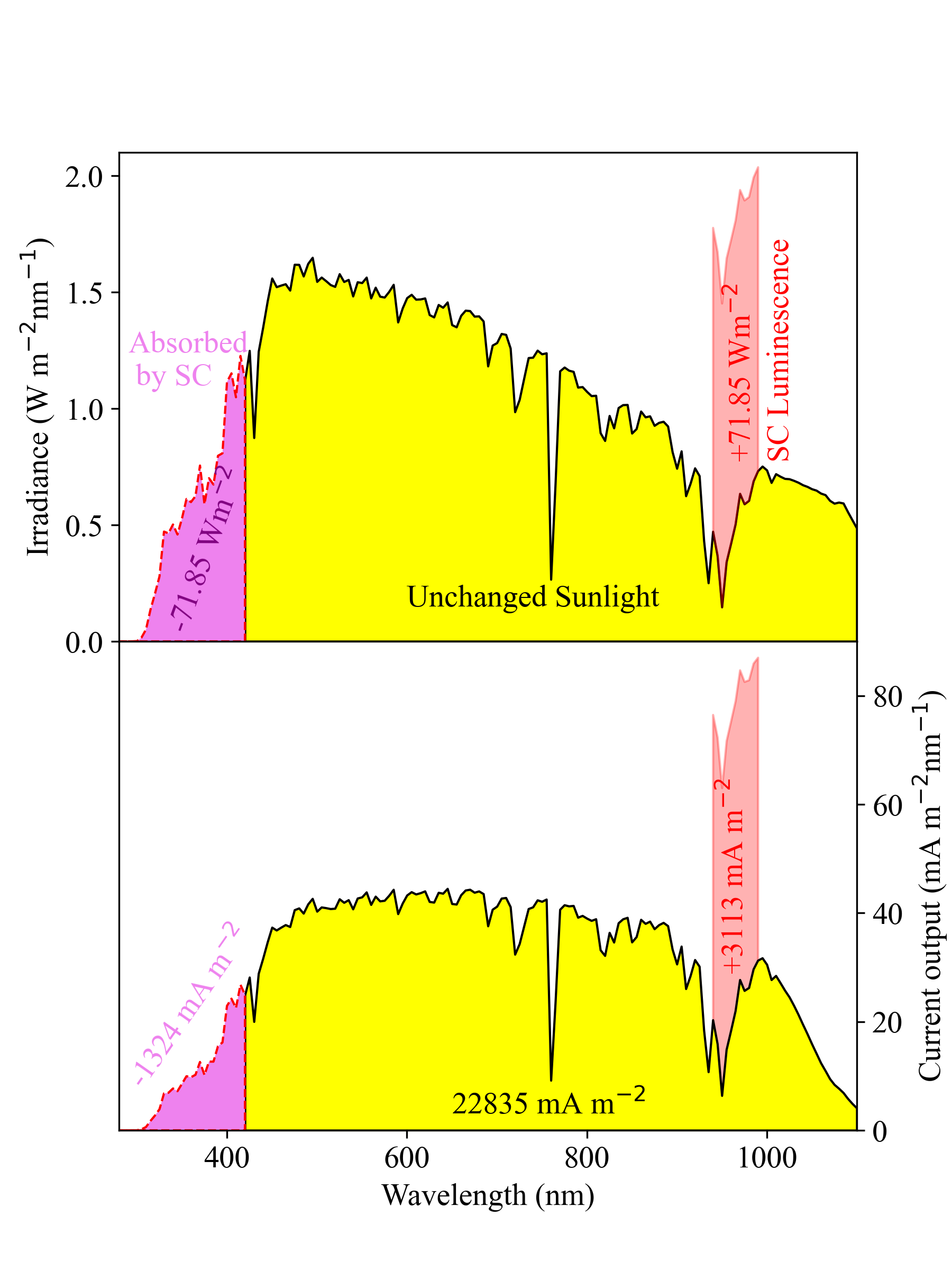}
    \caption{Ideal spectral converter (SC) effect on sunlight spectrum (top) and in the current output (bottom), demonstrating the balance between losses (at left, in purple) and the gains (at right, in red) that should remain positive for a viable SC.}
    \label{fig:SCideal}
\end{figure}

Using the reference spectra shown in Figure \ref{fig:scheme}, we could evaluate some estimates of this ideal SC effect on PV performance. The main results are presented in Figure \ref{fig:SCideal} and discussed next.

By integrating the irradiance spectrum below 420~nm we obtain a value of 71.85~Wm$^{-2}$ for the total radiation absorbed by the ideal SC. This radiation would produce 1324~mA~m$^{-2}$ if it could reach the PV, and we will call it ``SC current drop'' (C$_D$). Accordingly, it represents a primary adverse effect of the SC on the PV performance. As the total current output without the SC accounts for 24159~ mA~m$^{-2}$, even for our ideal SC, the current drop is significant, corresponding to 5.5~\% of the total. 

However, our ideal SC is supposed to have total efficiency in converting incident radiation. Assuming this same 71.85~Wm$^{-2}$ is emitted around 965~nm, it would produce an additional current output of about 3113~mA~m$^{-2}$, which we will call ``SC current gain'' (C$_G$). In other words, the net effect of a SC would be,
\begin{equation}
C_G-C_D=C_N,
\end{equation}
where $C_N$ is the net current for the SC. For reference, the proposed ideal SC described above would result in $C_N=1789~mA~m^{-2}$. This net current increment corresponds to a 7.4~\% increase in the total current output (from 24159 to 25948~mA~m$^{-2}$).

Unfortunately, collecting data from most of the materials proposed as SCs in the literature is challenging, and this review does not include a comparison between them in terms of the net current described above. We believe, though, this balance of negative and positive impacts of the SCs is critical, and its evaluation under our proposed framework can accelerate the development of SCs and be helpful to all researchers in this field.

In summary, our framework to characterize and compare materials proposed as SCs can be split into the following steps:
\begin{enumerate}
    \item Measure the material absorption spectrum;
    \item Define a thickness, depending on how this material is intended to be deployed in PV (nanometer/micrometer film, thick sample, etc);
    \item Calculate the absorbance spectrum for this specific material and thickness.
    \item Calculate C$_D$, due to this absorbance spectrum;
    \item Estimate the C$_G$, due to SC luminescence;
    \item Obtain the C$_N$ associated to the SC.
\end{enumerate}
We intentionally used the term "estimate" for the fifth step, recognizing the complexity of measuring or calculating luminescence. Even though luminescence can be quantified for one sample in a specific range, measuring it from the visible to the infrared often requires different sensors, gratings, etc. It may be challenging to have calibrated equipment to perform luminescence measurements in such a broad range. Comparing the luminescence intensity between several samples can be even more difficult, as reflection and scattering of the pump source and luminescent light self-absorption may vary among them (mainly for SCs containing resonant emitters, such as Yb$^{3+}$).

\section{Emerging trends}

\subsection{Lowering carbon emission with Hydrogen}
Scientists are pursuing to reduce $CO_2$ emissions related to glass production by replacing natural gas used in the float process. One of the most promising approaches is replacing natural gas with hydrogen, though this also looks challenging. The energy content of these two gases is significantly different, and the combustion of each one results in different amounts of heat transfer through radiation. The modified atmosphere composition inside the furnace also seems to increase the water and NO$_x$ content. Consequently, recent studies have addressed the effect of this hydrogen-modified float process on the optical quality of the final glass sheet~\cite{zier2021,griffin2021,furszyfer2022}.

By one hand it seems interesting to look for alternatives to the SLS glass, or modify it. However, by another, it seems likely that the float process itself will need adjustments to become greener. In both cases, there are plenty of work for the glass science community in understanding and proposing methods to optimize the quality of the float glass in light of this new challenge to fit the industry inside environmental boundaries.

\subsection{Alternative materials and methods}
\label{sec:multi}
A modern float SLS industry produces several thousand square meters of glass per hour~\cite{Belancon2021}, and some thin films may be produced inline with the process by chemical vapor deposition (CVD). Though this enables fast and cheap production of coatings, many other cannot be produced by the CVD method. In such a context, here we highlight some recent studies concerning materials, often produced by alternative methods, which are proposed for PV applications. 

Besides several glassy films~\cite{savi2022,Flores2019,Bubli2020} and glass-polymer matrices~\cite{Sandrini2023,Bouajaj2016,Enrichi2021}, one may found several innovative methods to introduce dopants~\cite{Muniz2021b} and coatings~\cite{Langenhorst2019c,Wang2022} which include even the use of organic molecules such as Chlorophyl~\cite{Wang2022}. As one can see, even though there are robust materials and techniques available at the industry level, the need to decarbonize the industry~\cite{zier2021,Bataille2018,furszyfer2022}, expand PV lifespan and efficiency as well as increase the reuse~\cite{Belancon2020} and recyclability~\cite{Lunardi2018,Ren2021,Dias2016a} of glass indicates a wide range of possibilities to be explored by glass scientists.

\section{Discussion}

Glass is undoubtedly an essential part of PV devices, and there is room for glass-related breakthroughs that could result in expanded net energy production of silicon based solar electricity. There is the possibility to develop CGs with reduced energy intensity and the need to reduce emissions from the flat glass production process.

On the other hand, particular features, such as self-cleaning and anti-reflective coatings, are already available at the industry level. Others, such as SCs still in the early research and development stage. Regarding this latter one, there are countless works published in the last few decades, though the different techniques and methods used to characterize SC materials make it quite hard to compare them.

Based on this fact, we proposed a benchmark framework for SCs, that could also be applied to UV filters for PV. As we have demonstrated, even narrow absorption bands can significantly affect the current output, mainly if those bands are between 450-900~nm. Our framework considers the ASTM-G173-03 sunlight spectrum and a commercial Si sensor response, resulting in a ``current output spectrum'' more suitable for comparing spectrally selective materials. Such a model seems straightforward and effective. If we aim to use a material to cover PVs, one should consider the spectral response of silicon and the overall net effect of this material in the electrical output.

As we have discussed in a previous work~\cite{Belancon2021}, a class AAA solar simulator should reproduce the solar irradiance within a margin of $\pm$25\% in six bands of the spectrum (five 100 nm wide and the last one between 900 and 1100 nm), using the AM1.5G as reference. Though such an approach is fundamental at the industry level to evaluate PVs, this low-resolution analysis is inadequate to compare and develop spectrally selective materials, namely SCs and UV filters. As demonstrated here, even a narrow absorption band just 5~nm wide of a cover material can significantly reduce the PV current output. The proposed model is an essential step towards a standard for theoretically and experimentally comparing different SCs.

Considering some of the emerging trends presented in this review, it seems clear that the glass industry is again looking for a revolution. If, on the one hand, there are not yet a candidate glass to replace the SLS in PVs, the need to decarbonize the glass industry by developing a fossil fuel-free float glass process is itself a breakthrough. This could also be an opportunity to develop a modified-SLS glass more suitable to this new environmentally friendly process or even to develop new glass systems. 

In conclusion, continued research and development in the field of spectrally selective materials could lead to significant advancements in the efficiency and sustainability of PV technology, ultimately contributing to a cleaner and greener future.

\section{Conclusion}

In this work, the literature on cover glasses and spectral converters has been reviewed. Several new glasses, glass ceramics, and multi-functional thins films have been investigated for PV applications in the last few years, and promising results have been reported. However, the quantitative comparison between these materials has not been performed correctly. There is a lack of standardized parameters to enable scientists to do so, which we propose here, as an instructive model for standardization.

Considering the AM1.5G solar spectrum and the Si's spectral response, our model enables a path to quantify the effect of these new materials on PV's electrical output, no matter where the absorption and emission bands are located. By calculating both the current drop and gain due to the SC, we can theoretically evaluate the effect of the spectral properties of the material on the electrical output. There are no restrictions to the material's geometry, so the model can also compare bulk and thin film materials.

The solar photovoltaic industry remains focused on Silicon technology. There are predictions of a critical increment in the share of bifacial solar panels in the following decades, evidencing we can expect an increment in flat glass demand for this sector. In this context, innovation is needed as it is mandatory to decarbonize the glass industry as soon as possible. Such a challenging task may be accomplished by replacing natural gas in the float process and developing modified SLS glasses. Besides that, expanding the efficiency or the lifespan of PVs will also contribute to reducing the environmental impact and, consequently, the cost of solar power worldwide.

This contribution summarizes the role of the cover glass in PVs, highlighting some of the most recent and exciting research results of glassy materials for solar silicon photovoltaic applications. The glass community has plenty of opportunities to develop new materials and processes that may reduce our carbon emissions and environmental footprint.

\section{Acknowledgement}

The authors would like to thank the \textit{Conselho Nacional de Desenvolvimento Científico e Tecnológico} (CNPq), Brazil, grant 409475/2021-1, and the Central de Análises laboratory in UTFPR-PB.


  \bibliographystyle{elsarticle-num} 

\begin{thebibliography}{100}
\expandafter\ifx\csname url\endcsname\relax
  \def\url#1{\texttt{#1}}\fi
\expandafter\ifx\csname urlprefix\endcsname\relax\def\urlprefix{URL }\fi
\expandafter\ifx\csname href\endcsname\relax
  \def\href#1#2{#2} \def\path#1{#1}\fi

\bibitem{Westbroek2021}
C.~D. Westbroek, J.~Bitting, M.~Craglia, J.~M.~C. Azevedo, J.~M. Cullen,
  \href{https://onlinelibrary.wiley.com/doi/10.1111/jiec.13112}{{Global
  material flow analysis of glass: From raw materials to end of life}}, Journal
  of Industrial Ecology 25~(2) (2021) 333--343.
\newblock \href {https://doi.org/10.1111/jiec.13112}
  {\path{doi:10.1111/jiec.13112}}.
\newline\urlprefix\url{https://onlinelibrary.wiley.com/doi/10.1111/jiec.13112}

\bibitem{Haegel2017}
N.~M. Haegel, R.~Margolis, T.~Buonassisi, D.~Feldman, A.~Froitzheim,
  R.~Garabedian, M.~Green, S.~Glunz, H.-m. Henning, B.~Holder, I.~Kaizuka,
  B.~Kroposki, K.~Matsubara, S.~Niki, K.~Sakurai, R.~A. Schindler, W.~Tumas,
  E.~R. Weber, G.~Wilson, M.~Woodhouse, S.~Kurtz,
  \href{http://www.sciencemag.org/lookup/doi/10.1126/science.aal1288}{{Terawatt-scale
  photovoltaics: Trajectories and challenges}}, Science 356~(6334) (2017)
  141--143.
\newblock \href {https://doi.org/10.1126/science.aal1288}
  {\path{doi:10.1126/science.aal1288}}.
\newline\urlprefix\url{http://www.sciencemag.org/lookup/doi/10.1126/science.aal1288}

\bibitem{Verlinden2020}
P.~J. Verlinden, {Future challenges for photovoltaic manufacturing at the
  terawatt level}, Journal of Renewable and Sustainable Energy 12~(5) (2020).
\newblock \href {https://doi.org/10.1063/5.0020380}
  {\path{doi:10.1063/5.0020380}}.

\bibitem{ITRPV2022}
{International Technology Roadmap for Photovoltaic (ITRPV) 2022}, Tech. rep.
  (2022).

\bibitem{sinha2021}
A.~Sinha, D.~B. Sulas-Kern, M.~Owen-Bellini, L.~Spinella,
  S.~Uli{\v{c}}n{\'{a}}, S.~{Ayala Pelaez}, S.~Johnston, L.~T. Schelhas,
  \href{https://iopscience.iop.org/article/10.1088/1361-6463/ac1462}{{Glass/glass
  photovoltaic module reliability and degradation: a review}}, Journal of
  Physics D: Applied Physics 54~(41) (2021) 413002.
\newblock \href {https://doi.org/10.1088/1361-6463/ac1462}
  {\path{doi:10.1088/1361-6463/ac1462}}.
\newline\urlprefix\url{https://iopscience.iop.org/article/10.1088/1361-6463/ac1462}

\bibitem{Haegel2019}
N.~M. Haegel, H.~Atwater, T.~Barnes, C.~Breyer, A.~Burrell, Y.-M. Chiang,
  S.~{De Wolf}, B.~Dimmler, D.~Feldman, S.~Glunz, J.~C. Goldschmidt,
  D.~Hochschild, R.~Inzunza, I.~Kaizuka, B.~Kroposki, S.~Kurtz, S.~Leu,
  R.~Margolis, K.~Matsubara, A.~Metz, W.~K. Metzger, M.~Morjaria, S.~Niki,
  S.~Nowak, I.~M. Peters, S.~Philipps, T.~Reindl, A.~Richter, D.~Rose,
  K.~Sakurai, R.~Schlatmann, M.~Shikano, W.~Sinke, R.~Sinton, B.~Stanbery,
  M.~Topic, W.~Tumas, Y.~Ueda, J.~van~de Lagemaat, P.~Verlinden, M.~Vetter,
  E.~Warren, M.~Werner, M.~Yamaguchi, A.~W. Bett, {Terawatt-scale
  photovoltaics: Transform global energy}, Science 364~(6443) (2019) 836--838.
\newblock \href {https://doi.org/10.1126/science.aaw1845}
  {\path{doi:10.1126/science.aaw1845}}.

\bibitem{Davidsson2016}
S.~Davidsson, M.~H{\"{o}}{\"{o}}k,
  \href{http://dx.doi.org/10.1016/j.enpol.2017.06.028}{{Material requirements
  and availability for multi-terawatt deployment of photovoltaics}}, Energy
  Policy 108~(June) (2016) 574--582.
\newblock \href {https://doi.org/10.1016/j.enpol.2017.06.028}
  {\path{doi:10.1016/j.enpol.2017.06.028}}.
\newline\urlprefix\url{http://dx.doi.org/10.1016/j.enpol.2017.06.028}

\bibitem{Tao2011}
C.~S. Tao, J.~Jiang, M.~Tao,
  \href{http://dx.doi.org/10.1016/j.solmat.2011.06.013}{{Natural resource
  limitations to terawatt-scale solar cells}}, Solar Energy Materials and Solar
  Cells 95~(12) (2011) 3176--3180.
\newblock \href {https://doi.org/10.1016/j.solmat.2011.06.013}
  {\path{doi:10.1016/j.solmat.2011.06.013}}.
\newline\urlprefix\url{http://dx.doi.org/10.1016/j.solmat.2011.06.013}

\bibitem{Davidsson2017}
S.~Davidsson, M.~H{\"{o}}{\"{o}}k,
  \href{https://linkinghub.elsevier.com/retrieve/pii/S0301421517303798}{{Material
  requirements and availability for multi-terawatt deployment of
  photovoltaics}}, Energy Policy 108~(December 2016) (2017) 574--582.
\newblock \href {https://doi.org/10.1016/j.enpol.2017.06.028}
  {\path{doi:10.1016/j.enpol.2017.06.028}}.
\newline\urlprefix\url{https://linkinghub.elsevier.com/retrieve/pii/S0301421517303798}

\bibitem{Bazilian2018}
M.~D. Bazilian, \href{http://dx.doi.org/10.1016/j.exis.2017.12.002}{{The
  mineral foundation of the energy transition}}, Extractive Industries and
  Society 5~(1) (2018) 93--97.
\newblock \href {https://doi.org/10.1016/j.exis.2017.12.002}
  {\path{doi:10.1016/j.exis.2017.12.002}}.
\newline\urlprefix\url{http://dx.doi.org/10.1016/j.exis.2017.12.002}

\bibitem{Burrows2015}
K.~Burrows, V.~Fthenakis,
  \href{http://dx.doi.org/10.1016/j.solmat.2014.09.028}{{Glass needs for a
  growing photovoltaics industry}}, Solar Energy Materials and Solar Cells 132
  (2015) 455--459.
\newblock \href {https://doi.org/10.1016/j.solmat.2014.09.028}
  {\path{doi:10.1016/j.solmat.2014.09.028}}.
\newline\urlprefix\url{http://dx.doi.org/10.1016/j.solmat.2014.09.028}

\bibitem{FurszyferDelRio2022}
D.~D. {Furszyfer Del Rio}, B.~K. Sovacool, A.~M. Foley, S.~Griffiths,
  M.~Bazilian, J.~Kim, D.~Rooney,
  \href{https://doi.org/10.1016/j.rser.2021.111885}{{Decarbonizing the glass
  industry: A critical and systematic review of developments, sociotechnical
  systems and policy options}}, Renewable and Sustainable Energy Reviews
  155~(November 2021) (2022) 111885.
\newblock \href {https://doi.org/10.1016/j.rser.2021.111885}
  {\path{doi:10.1016/j.rser.2021.111885}}.
\newline\urlprefix\url{https://doi.org/10.1016/j.rser.2021.111885}

\bibitem{Allsopp2020}
B.~L. Allsopp, R.~Orman, S.~R. Johnson, I.~Baistow, G.~Sanderson, P.~Sundberg,
  C.~St{\aa}lhandske, L.~Grund, A.~Andersson, J.~Booth, P.~A. Bingham,
  S.~Karlsson, {Towards improved cover glasses for photovoltaic devices},
  Progress in Photovoltaics: Research and Applications 28~(11) (2020)
  1187--1206.
\newblock \href {https://doi.org/10.1002/pip.3334}
  {\path{doi:10.1002/pip.3334}}.

\bibitem{Shockley1961}
W.~Shockley, H.~J. Queisser, {Detailed balance limit of efficiency of p-n
  junction solar cells}, Journal of Applied Physics 32~(3) (1961) 510--519.
\newblock \href {https://doi.org/10.1063/1.1736034}
  {\path{doi:10.1063/1.1736034}}.

\bibitem{Guillemoles2019}
J.-F. Guillemoles, T.~Kirchartz, D.~Cahen, U.~Rau,
  \href{http://www.nature.com/articles/s41566-019-0479-2}{{Guide for the
  perplexed to the Shockley–Queisser model for solar cells}}, Nature
  Photonics 13~(8) (2019) 501--505.
\newblock \href {https://doi.org/10.1038/s41566-019-0479-2}
  {\path{doi:10.1038/s41566-019-0479-2}}.
\newline\urlprefix\url{http://www.nature.com/articles/s41566-019-0479-2}

\bibitem{Hall2014}
C.~A.~S. Hall, J.~G. Lambert, S.~B. Balogh,
  \href{http://dx.doi.org/10.1016/j.enpol.2013.05.049}{{EROI of different fuels
  and the implications for society}}, Energy Policy 64 (2014) 141--152.
\newblock \href {https://doi.org/10.1016/j.enpol.2013.05.049}
  {\path{doi:10.1016/j.enpol.2013.05.049}}.
\newline\urlprefix\url{http://dx.doi.org/10.1016/j.enpol.2013.05.049}

\bibitem{Bhandari2015}
K.~P. Bhandari, J.~M. Collier, R.~J. Ellingson, D.~S. Apul,
  \href{http://dx.doi.org/10.1016/j.rser.2015.02.057}{{Energy payback time
  (EPBT) and energy return on energy invested (EROI) of solar photovoltaic
  systems: A systematic review and meta-analysis}}, Renewable and Sustainable
  Energy Reviews 47 (2015) 133--141.
\newblock \href {https://doi.org/10.1016/j.rser.2015.02.057}
  {\path{doi:10.1016/j.rser.2015.02.057}}.
\newline\urlprefix\url{http://dx.doi.org/10.1016/j.rser.2015.02.057}

\bibitem{bhandary2015}
K.~P. Bhandari, J.~M. Collier, R.~J. Ellingson, D.~S. Apul, {Energy payback
  time (EPBT) and energy return on energy invested (EROI) of solar photovoltaic
  systems: A systematic review and meta-analysis} (2015).
\newblock \href {https://doi.org/10.1016/j.rser.2015.02.057}
  {\path{doi:10.1016/j.rser.2015.02.057}}.

\bibitem{diesendorf2020}
M.~Diesendorf, T.~Wiedmann, {Implications of Trends in Energy Return on Energy
  Invested (EROI) for Transitioning to Renewable Electricity}, Ecological
  Economics 176 (oct 2020).
\newblock \href {https://doi.org/10.1016/j.ecolecon.2020.106726}
  {\path{doi:10.1016/j.ecolecon.2020.106726}}.

\bibitem{Zhou2018}
Z.~Zhou, M.~Carbajales-Dale, {Assessing the photovoltaic technology landscape:
  Efficiency and energy return on investment (EROI)}, Energy and Environmental
  Science 11~(3) (2018) 603--608.
\newblock \href {https://doi.org/10.1039/c7ee01806a}
  {\path{doi:10.1039/c7ee01806a}}.

\bibitem{Pilkington1969}
L.~A.~B. Pilkington,
  \href{https://royalsocietypublishing.org/doi/10.1098/rspa.1969.0212}{{Review
  Lecture: The float glass process}}, Proceedings of the Royal Society of
  London. A. Mathematical and Physical Sciences 314~(1516) (1969) 1--25.
\newblock \href {https://doi.org/10.1098/rspa.1969.0212}
  {\path{doi:10.1098/rspa.1969.0212}}.
\newline\urlprefix\url{https://royalsocietypublishing.org/doi/10.1098/rspa.1969.0212}

\bibitem{nascimento2014}
M.~L.~F. Nascimento,
  \href{https://linkinghub.elsevier.com/retrieve/pii/S0172219014000507}{{Brief
  history of the flat glass patent – Sixty years of the float process}},
  World Patent Information 38 (2014) 50--56.
\newblock \href {https://doi.org/10.1016/j.wpi.2014.04.006}
  {\path{doi:10.1016/j.wpi.2014.04.006}}.
\newline\urlprefix\url{https://linkinghub.elsevier.com/retrieve/pii/S0172219014000507}

\bibitem{youngman2021}
R.~E. Youngman,
  \href{https://onlinelibrary.wiley.com/doi/10.1002/9781118801017.ch7.6}{{Borosilicate
  Glasses}}, in: Encyclopedia of Glass Science, Technology, History, and
  Culture, Vol.~35, Wiley, 2021, pp. 867--878.
\newblock \href {https://doi.org/10.1002/9781118801017.ch7.6}
  {\path{doi:10.1002/9781118801017.ch7.6}}.
\newline\urlprefix\url{https://onlinelibrary.wiley.com/doi/10.1002/9781118801017.ch7.6}

\bibitem{manara2009}
D.~Manara, A.~Grandjean, D.~R. Neuville, {Structure of borosilicate glasses and
  melts: A revision of the Yun, Bray and Dell model}, Journal of
  Non-Crystalline Solids 355~(50-51) (2009) 2528--2531.
\newblock \href {https://doi.org/10.1016/j.jnoncrysol.2009.08.033}
  {\path{doi:10.1016/j.jnoncrysol.2009.08.033}}.

\bibitem{angeli2012}
F.~Angeli, O.~Villain, S.~Schuller, T.~Charpentier, D.~{De Ligny}, L.~Bressel,
  L.~Wondraczek, {Effect of temperature and thermal history on borosilicate
  glass structure}, Physical Review B - Condensed Matter and Materials Physics
  85~(5) (feb 2012).
\newblock \href {https://doi.org/10.1103/PhysRevB.85.054110}
  {\path{doi:10.1103/PhysRevB.85.054110}}.

\bibitem{tostanoski2022}
N.~J. Tostanoski, D.~M{\"{o}}ncke, R.~Youngman, S.~K. Sundaram,
  {Structure-terahertz property relationship in sodium borosilicate glasses},
  International Journal of Applied Glass Science (2022).
\newblock \href {https://doi.org/10.1111/ijag.16608}
  {\path{doi:10.1111/ijag.16608}}.

\bibitem{perea2020}
D.~E. Perea, D.~K. Schreiber, J.~V. Ryan, M.~G. Wirth, L.~Deng, X.~Lu, J.~Du,
  J.~D. Vienna, {Tomographic mapping of the nanoscale water-filled pore
  structure in corroded borosilicate glass}, npj Materials Degradation 4~(1)
  (dec 2020).
\newblock \href {https://doi.org/10.1038/s41529-020-0110-5}
  {\path{doi:10.1038/s41529-020-0110-5}}.

\bibitem{zhu2019}
H.~Zhu, F.~Wang, Q.~Liao, D.~Liu, Y.~Zhu, {Structure features, crystallization
  kinetics and water resistance of borosilicate glasses doped with CeO2},
  Journal of Non-Crystalline Solids 518 (2019) 57--65.
\newblock \href {https://doi.org/10.1016/j.jnoncrysol.2019.04.044}
  {\path{doi:10.1016/j.jnoncrysol.2019.04.044}}.

\bibitem{Guyot2011a}
Y.~Guyot, A.~Steimacher, M.~P.~M. Belan{\c{c}}on, A.~A.~N. Medina, M.~M.~L.
  Baesso, S.~S.~M. Lima, L.~L. H.~C. Andrade, A.~Brenier, A.-M. A.-M. Jurdyc,
  G.~Boulon,
  \href{https://www.osapublishing.org/abstract.cfm?URI=josab-28-10-2510
  https://opg.optica.org/abstract.cfm?URI=josab-28-10-2510}{{Spectroscopic
  properties, concentration quenching, and laser investigations of
  Yb$^{3+}$-doped calcium aluminosilicate glasses}}, Journal of the Optical
  Society of America B 28~(10) (2011) 2510.
\newblock \href {https://doi.org/10.1364/JOSAB.28.002510}
  {\path{doi:10.1364/JOSAB.28.002510}}.
\newline\urlprefix\url{https://www.osapublishing.org/abstract.cfm?URI=josab-28-10-2510
  https://opg.optica.org/abstract.cfm?URI=josab-28-10-2510}

\bibitem{belancon2014}
M.~P. Belan{\c{c}}on, J.~D. Marconi, M.~F. Ando, L.~C. Barbosa, {Near-IR
  emission in {P}{r$^{3+}$} single doped and tunable near-IR emission in
  {P}{$r^{3+}$}/{Y}{$b^{3+}$} codoped tellurite tungstate glasses for broadband
  optical amplifiers}, Optical Materials 36~(6) (2014) 1020--1026.

\bibitem{Dragic2012b}
P.~Dragic, T.~Hawkins, P.~Foy, S.~Morris, J.~Ballato,
  \href{http://dx.doi.org/10.1038/nphoton.2012.182}{{Sapphire-derived all-glass
  optical fibres}}, Nature Photonics 6~(9) (2012) 629--635.
\newblock \href {https://doi.org/10.1038/nphoton.2012.182}
  {\path{doi:10.1038/nphoton.2012.182}}.
\newline\urlprefix\url{http://dx.doi.org/10.1038/nphoton.2012.182}

\bibitem{Belancon2013d}
M.~P. Belan{\c{c}}on, M.~F. Ando, J.~D. Marconi, H.~N. Yoshimura, E.~F.
  Chillcce, L.~C. Barbosa, H.~L. Fragnito,
  \href{https://opg.optica.org/abstract.cfm?URI=WSOF-2013-F2.25}{{Tellurite
  microstructured optical fibers doped with rare-earths for optical
  amplification}}, in: Workshop on Specialty Optical Fibers and their
  Applications, OSA, Washington, D.C., 2013, p. F2.25.
\newblock \href {https://doi.org/10.1364/WSOF.2013.F2.25}
  {\path{doi:10.1364/WSOF.2013.F2.25}}.
\newline\urlprefix\url{https://opg.optica.org/abstract.cfm?URI=WSOF-2013-F2.25}

\bibitem{Zhong2018}
J.~Zhong, D.~Chen, Y.~Peng, Y.~Lu, X.~Chen, X.~Li, Z.~Ji,
  \href{https://doi.org/10.1016/j.jallcom.2018.05.348}{{A review on
  nanostructured glass ceramics for promising application in optical
  thermometry}}, Journal of Alloys and Compounds 763 (2018) 34--48.
\newblock \href {https://doi.org/10.1016/j.jallcom.2018.05.348}
  {\path{doi:10.1016/j.jallcom.2018.05.348}}.
\newline\urlprefix\url{https://doi.org/10.1016/j.jallcom.2018.05.348}

\bibitem{Du2019}
B.~Du, Z.~Tong, X.~Mu, J.~Xu, S.~Liu, Z.~Liu, W.~Cao, Z.~M. Qi, {A potassium
  ion-exchanged glass opticalwaveguide sensor locally coated with a crystal
  violet-sio2 gel film for real-time detection of organophosphorus pesticides
  simulant}, Sensors (Switzerland) 19~(19) (2019) 1--12.
\newblock \href {https://doi.org/10.3390/s19194219}
  {\path{doi:10.3390/s19194219}}.

\bibitem{Chen2020a}
S.~Y.~Z. Chen, W.~H. Song, J.~K. Cao, F.~F. Hu, H.~Guo,
  \href{https://doi.org/10.1016/j.jallcom.2020.154011}{{Highly sensitive
  optical thermometer based on FIR technique of transparent NaY2F7:Tm3+/Yb3+
  glass ceramic}}, Journal of Alloys and Compounds 825 (2020) 154011.
\newblock \href {https://doi.org/10.1016/j.jallcom.2020.154011}
  {\path{doi:10.1016/j.jallcom.2020.154011}}.
\newline\urlprefix\url{https://doi.org/10.1016/j.jallcom.2020.154011}

\bibitem{Sandrini2018}
M.~Sandrini, R.~F. Muniz, V.~S. Zanuto, F.~Pedrochi, Y.~Guyot, A.~C. Bento,
  M.~L. Baesso, A.~Steimacher, A.~M. Neto, {Enhanced and tunable white light
  emission from Ag nanoclusters and Eu3+-co-doped CaBAl glasses}, RSC Advances
  8~(61) (2018) 35263--35270.
\newblock \href {https://doi.org/10.1039/c8ra07114d}
  {\path{doi:10.1039/c8ra07114d}}.

\bibitem{Chen2014}
L.-Y. Chen, W.-C. Cheng, C.-C. Tsai, Y.-C. Huang, Y.-S. Lin, W.-H. Cheng,
  \href{https://www.osapublishing.org/ome/abstract.cfm?uri=ome-4-1-121}{{High-performance
  glass phosphor for white-light-emitting diodes via reduction of Si-Ce$^{3+}
  $:YAG inter-diffusion}}, Optical Materials Express 4~(1) (2014) 121.
\newblock \href {https://doi.org/10.1364/OME.4.000121}
  {\path{doi:10.1364/OME.4.000121}}.
\newline\urlprefix\url{https://www.osapublishing.org/ome/abstract.cfm?uri=ome-4-1-121}

\bibitem{Wang2020}
H.~Wang, Y.~Mou, Y.~Peng, Y.~Zhang, A.~Wang, L.~Xu, H.~Long, M.~Chen, J.~Dai,
  C.~Chen, \href{https://doi.org/10.1016/j.jallcom.2019.152321}{{Fabrication of
  phosphor glass film on aluminum plate by using lead-free tellurite glass for
  laser-driven white lighting}}, Journal of Alloys and Compounds 814 (2020)
  152321.
\newblock \href {https://doi.org/10.1016/j.jallcom.2019.152321}
  {\path{doi:10.1016/j.jallcom.2019.152321}}.
\newline\urlprefix\url{https://doi.org/10.1016/j.jallcom.2019.152321}

\bibitem{Kang2008}
J.~G. Kang, M.~K. Kim, K.~B. Kim,
  \href{http://linkinghub.elsevier.com/retrieve/pii/S0025540807004400}{{Preparation
  and luminescence characterization of GGAG:Ce3+,B3+ for a white light-emitting
  diode}}, Materials Research Bulletin 43~(8-9) (2008) 1982--1988.
\newblock \href {https://doi.org/10.1016/j.materresbull.2007.10.001}
  {\path{doi:10.1016/j.materresbull.2007.10.001}}.
\newline\urlprefix\url{http://linkinghub.elsevier.com/retrieve/pii/S0025540807004400}

\bibitem{Silveira2012a}
L.~G. Silveira, L.~F. C{\'{o}}tica, I.~A. Santos, M.~P. Belan{\c{c}}on, J.~H.
  Rohling, M.~L. Baesso,
  \href{http://linkinghub.elsevier.com/retrieve/pii/S0167577X12012268
  https://linkinghub.elsevier.com/retrieve/pii/S0167577X12012268}{{Processing
  and luminescence properties of Ce:Y3Al5O12and Eu:Y3Al5O12ceramics for
  white-light applications}}, Materials Letters 89 (2012) 86--89.
\newblock \href {https://doi.org/10.1016/j.matlet.2012.08.106}
  {\path{doi:10.1016/j.matlet.2012.08.106}}.
\newline\urlprefix\url{http://linkinghub.elsevier.com/retrieve/pii/S0167577X12012268
  https://linkinghub.elsevier.com/retrieve/pii/S0167577X12012268}

\bibitem{Yi2015}
S.~Yi, W.~J. Chung, J.~Heo,
  \href{http://dx.doi.org/10.1186/s40539-015-0027-8}{{Phosphor-in-glasses
  composites containing light diffusers for high color uniformity of
  white-light-emitting diodes}}, Journal of Solid State Lighting 2~(1) (2015)
  4--9.
\newblock \href {https://doi.org/10.1186/s40539-015-0027-8}
  {\path{doi:10.1186/s40539-015-0027-8}}.
\newline\urlprefix\url{http://dx.doi.org/10.1186/s40539-015-0027-8}

\bibitem{Zhu2017}
Q.~Q. Zhu, X.~Xu, L.~Wang, Z.~F. Tian, Y.~Z. Xu, N.~Hirosaki, R.~J. Xie,
  \href{http://dx.doi.org/10.1016/j.jallcom.2017.01.256}{{A robust red-emitting
  phosphor-in-glass (PiG) for use in white lighting sources pumped by blue
  laser diodes}}, Journal of Alloys and Compounds 702 (2017) 193--198.
\newblock \href {https://doi.org/10.1016/j.jallcom.2017.01.256}
  {\path{doi:10.1016/j.jallcom.2017.01.256}}.
\newline\urlprefix\url{http://dx.doi.org/10.1016/j.jallcom.2017.01.256}

\bibitem{Zheng2018}
P.~Zheng, S.~Li, L.~Wang, T.~L. Zhou, S.~You, T.~Takeda, N.~Hirosaki, R.~J.
  Xie, {Unique Color Converter Architecture Enabling Phosphor-in-Glass (PiG)
  Films Suitable for High-Power and High-Luminance Laser-Driven White
  Lighting}, ACS Applied Materials and Interfaces 10~(17) (2018) 14930--14940.
\newblock \href {https://doi.org/10.1021/acsami.8b03168}
  {\path{doi:10.1021/acsami.8b03168}}.

\bibitem{sandrini2020}
M.~Sandrini, R.~F. Muniz, V.~S. Zanuto, J.~R. Marques~Viana, R.~D. Bini, J.~H.
  Rohling, M.~L. Baesso, A.~Medina~Neto,
  \href{https://linkinghub.elsevier.com/retrieve/pii/S0925838819338976}{{Glass
  frits as an enabler in the production process of OH$^{-}$-free calcium
  aluminosilicate glasses}}, Journal of Alloys and Compounds 816 (2020) 152651.
\newblock \href {https://doi.org/10.1016/j.jallcom.2019.152651}
  {\path{doi:10.1016/j.jallcom.2019.152651}}.
\newline\urlprefix\url{https://linkinghub.elsevier.com/retrieve/pii/S0925838819338976}

\bibitem{savi2022}
E.~L. Savi, R.~F. Muniz, A.~A.~J. Silva, G.~J. Schiavon, J.~W. Berrar, F.~R.
  Estrada, P.~Schio, J.~{C. Cezar}, J.~H. Rohling, V.~S. Zanuto, A.~C. Bento,
  A.~N. Medina, L.~A.~O. Nunes, M.~L. Baesso,
  \href{https://aip.scitation.org/doi/10.1063/5.0067794}{{Thin-film of Nd 3+
  –Yb 3+ co-doped low silica calcium aluminosilicate glass grown by a laser
  deposition technique}}, Journal of Applied Physics 131~(5) (2022) 055304.
\newblock \href {https://doi.org/10.1063/5.0067794}
  {\path{doi:10.1063/5.0067794}}.
\newline\urlprefix\url{https://aip.scitation.org/doi/10.1063/5.0067794}

\bibitem{Tai2018}
Y.~Tai, X.~Li, B.~Pan,
  \href{https://doi.org/10.1016/j.jlumin.2017.10.051}{{Efficient near-infrared
  down conversion in Nd3+-Yb3+ co-doped transparent nanostructured glass
  ceramics for photovoltaic application}}, Journal of Luminescence
  195~(September 2017) (2018) 102--108.
\newblock \href {https://doi.org/10.1016/j.jlumin.2017.10.051}
  {\path{doi:10.1016/j.jlumin.2017.10.051}}.
\newline\urlprefix\url{https://doi.org/10.1016/j.jlumin.2017.10.051}

\bibitem{Fu2012b}
H.~Fu, S.~Cui, Q.~Luo, X.~Qiao, X.~Fan, X.~Zhang,
  \href{http://dx.doi.org/10.1016/j.jnoncrysol.2012.02.024}{{Broadband
  downshifting luminescence of Cr3+/Yb3+-codoped fluorosilicate glass}},
  Journal of Non-Crystalline Solids 358~(9) (2012) 1217--1220.
\newblock \href {http://arxiv.org/abs/arXiv:1209.4226v1}
  {\path{arXiv:arXiv:1209.4226v1}}, \href
  {https://doi.org/10.1016/j.jnoncrysol.2012.02.024}
  {\path{doi:10.1016/j.jnoncrysol.2012.02.024}}.
\newline\urlprefix\url{http://dx.doi.org/10.1016/j.jnoncrysol.2012.02.024}

\bibitem{Bengtsson2022}
F.~Bengtsson, I.~B. Pehlivan, L.~{\"{O}}sterlund, S.~Karlsson, {Alkali ion
  diffusion and structure of chemically strengthened TiO2 doped soda-lime
  silicate glass}, Journal of Non-Crystalline Solids 586~(January) (2022).
\newblock \href {https://doi.org/10.1016/j.jnoncrysol.2022.121564}
  {\path{doi:10.1016/j.jnoncrysol.2022.121564}}.

\bibitem{Babkina2019}
A.~N. Babkina, K.~S. Zyryanova, D.~A. Agafonova, R.~K. Nuryev, A.~I. Ignatiev,
  D.~Valiev, \href{https://doi.org/10.1016/j.jnoncrysol.2019.119487}{{The
  effect of chromium concentration on luminescent properties of
  alkali-alumina-borate glass-ceramics}}, Journal of Non-Crystalline Solids
  521~(June) (2019) 119487.
\newblock \href {https://doi.org/10.1016/j.jnoncrysol.2019.119487}
  {\path{doi:10.1016/j.jnoncrysol.2019.119487}}.
\newline\urlprefix\url{https://doi.org/10.1016/j.jnoncrysol.2019.119487}

\bibitem{Muniz2021}
R.~F. Muniz, V.~O. Soares, G.~H. Montagnini, A.~N. Medina, M.~L. Baesso,
  \href{https://doi.org/10.1016/j.ceramint.2021.05.224}{{Thermal, optical and
  structural properties of relatively depolymerized sodium calcium silicate
  glass and glass-ceramic containing CaF2}}, Ceramics International 47~(17)
  (2021) 24966--24972.
\newblock \href {https://doi.org/10.1016/j.ceramint.2021.05.224}
  {\path{doi:10.1016/j.ceramint.2021.05.224}}.
\newline\urlprefix\url{https://doi.org/10.1016/j.ceramint.2021.05.224}

\bibitem{Muniz2021a}
R.~F. Muniz, A.~Steimacher, F.~Pedrochi, V.~S. Zanuto, L.~M. Azevedo, J.~H.
  Rohling, M.~L. Baesso, A.~N. Medina,
  \href{https://doi.org/10.1016/j.jallcom.2021.161484}{{Eu2+-Nd3+ co-doped
  glasses for solar spectrum modification via NUV/visible to NIR
  downconversion}}, Journal of Alloys and Compounds 888 (2021) 161484.
\newblock \href {https://doi.org/10.1016/j.jallcom.2021.161484}
  {\path{doi:10.1016/j.jallcom.2021.161484}}.
\newline\urlprefix\url{https://doi.org/10.1016/j.jallcom.2021.161484}

\bibitem{Gomez2011}
S.~G{\'{o}}mez, I.~Urra, R.~Valiente, F.~Rodr{\'{i}}guez,
  \href{https://linkinghub.elsevier.com/retrieve/pii/S0927024810004472}{{Spectroscopic
  study of Cu2+/Cu+ doubly doped and highly transmitting glasses for solar
  spectral transformation}}, Solar Energy Materials and Solar Cells 95~(8)
  (2011) 2018--2022.
\newblock \href {https://doi.org/10.1016/j.solmat.2010.07.022}
  {\path{doi:10.1016/j.solmat.2010.07.022}}.
\newline\urlprefix\url{https://linkinghub.elsevier.com/retrieve/pii/S0927024810004472}

\bibitem{Gomez-Salces2016}
S.~G{\'{o}}mez-Salces, J.~A. Barreda-Arg{\"{u}}eso, R.~Valiente,
  F.~Rodr{\'{i}}guez, {A study of Ce3+ to Mn2+ energy transfer in high
  transmission glasses using time-resolved spectroscopy}, Journal of Materials
  Chemistry C 4~(38) (2016) 9021--9026.
\newblock \href {https://doi.org/10.1039/c6tc01408a}
  {\path{doi:10.1039/c6tc01408a}}.

\bibitem{YULIANTINI}
L.~Yuliantini, N.~Nursam, L.~Pranoto, Shobih, J.~Hidayat, R.~Sova, Isnaeni,
  E.~Rahayu, M.~Djamal, P.~Yasaka, K.~Boonin, J.~Kaewkhao,
  \href{https://www.sciencedirect.com/science/article/pii/S0925838823014664}{Photon
  up-conversion in er3+ ion-doped zno-al2o3-bao-b2o3 glass for enhancing the
  performance of dye-sensitized solar cells}, Journal of Alloys and Compounds
  954 (2023) 170163.
\newblock \href {https://doi.org/https://doi.org/10.1016/j.jallcom.2023.170163}
  {\path{doi:https://doi.org/10.1016/j.jallcom.2023.170163}}.
\newline\urlprefix\url{https://www.sciencedirect.com/science/article/pii/S0925838823014664}

\bibitem{Reisfeld1972}
R.~Reisfeld, L.~Boehm,
  \href{https://linkinghub.elsevier.com/retrieve/pii/0022459672901570}{{Energy
  transfer between samarium and europium in phosphate glasses}}, Journal of
  Solid State Chemistry 4~(3) (1972) 417--424.
\newblock \href {https://doi.org/10.1016/0022-4596(72)90157-0}
  {\path{doi:10.1016/0022-4596(72)90157-0}}.
\newline\urlprefix\url{https://linkinghub.elsevier.com/retrieve/pii/0022459672901570}

\bibitem{REISFELD1978}
R.~REISFELD, S.~NEUMAN, \href{http://www.nature.com/articles/274144a0}{{Planar
  solar energy converter and concentrator based on uranyl-doped glass}}, Nature
  274~(5667) (1978) 144--145.
\newblock \href {https://doi.org/10.1038/274144a0}
  {\path{doi:10.1038/274144a0}}.
\newline\urlprefix\url{http://www.nature.com/articles/274144a0}

\bibitem{Reisfeld1980}
R.~Reisfeld, Y.~Kalisky,
  \href{http://www.nature.com/articles/283281a0}{{Improved planar solar
  converter based on uranyl neodymium and holmium glasses}}, Nature 283~(5744)
  (1980) 281--282.
\newblock \href {https://doi.org/10.1038/283281a0}
  {\path{doi:10.1038/283281a0}}.
\newline\urlprefix\url{http://www.nature.com/articles/283281a0}

\bibitem{Pan2000}
A.~Pan, A.~Ghosh, {New family of lead-bismuthate glass with a large
  transmitting window}, Journal of Non-Crystalline Solids 271~(1) (2000)
  157--161.
\newblock \href {https://doi.org/10.1016/S0022-3093(00)00111-3}
  {\path{doi:10.1016/S0022-3093(00)00111-3}}.

\bibitem{HumbertodaCunhaAndrade2018}
A.~K. {Rufino Souza}, A.~P. Langaro, J.~R. Silva, F.~B. Costa, K.~Yukimitu,
  J.~C. {Silos Moraes}, L.~{Antonio de Oliveira Nunes}, L.~{Humberto da Cunha
  Andrade}, S.~M. Lima,
  \href{https://linkinghub.elsevier.com/retrieve/pii/S0925838818345961}{{On the
  efficient Te4+-Yb3+ cooperative energy transfer mechanism in tellurite
  glasses: A potential material for luminescent solar concentrators}}, Journal
  of Alloys and Compounds 781 (2019) 1119--1126.
\newblock \href {https://doi.org/10.1016/j.jallcom.2018.12.038}
  {\path{doi:10.1016/j.jallcom.2018.12.038}}.
\newline\urlprefix\url{https://linkinghub.elsevier.com/retrieve/pii/S0925838818345961}

\bibitem{Leonard2013a}
R.~L. Leonard, S.~K. Gray, S.~D. Albritton, L.~N. Brothers, R.~M. Cross, A.~N.
  Eastes, H.~Y. Hah, H.~S. James, J.~E. King, S.~R. Mishra, J.~A. Johnson,
  \href{http://dx.doi.org/10.1016/j.jnoncrysol.2013.01.029}{{Rare earth doped
  downshifting glass ceramics for photovoltaic applications}}, Journal of
  Non-Crystalline Solids 366~(1) (2013) 1--5.
\newblock \href {https://doi.org/10.1016/j.jnoncrysol.2013.01.029}
  {\path{doi:10.1016/j.jnoncrysol.2013.01.029}}.
\newline\urlprefix\url{http://dx.doi.org/10.1016/j.jnoncrysol.2013.01.029}

\bibitem{Ahrens2011}
B.~Ahrens, S.~Brand, T.~B{\"{u}}chner, P.~Darr, S.~Schoenfelder,
  C.~Pa{\ss}lick, S.~Schweizer,
  \href{http://dx.doi.org/10.1016/j.jnoncrysol.2010.11.084}{{Mechanical
  properties of fluorozirconate-based glass ceramics for medical and
  photovoltaic applications}}, Journal of Non-Crystalline Solids 357~(11-13)
  (2011) 2264--2267.
\newblock \href {https://doi.org/10.1016/j.jnoncrysol.2010.11.084}
  {\path{doi:10.1016/j.jnoncrysol.2010.11.084}}.
\newline\urlprefix\url{http://dx.doi.org/10.1016/j.jnoncrysol.2010.11.084}

\bibitem{Taniguchi2019}
M.~M. Taniguchi, V.~S. Zanuto, P.~N. Portes, L.~C. Malacarne, N.~G.~C. Astrath,
  J.~D. Marconi, M.~P. Belan{\c{c}}on, \href{10.1016/j.jnoncrysol.2019.119717
  https://linkinghub.elsevier.com/retrieve/pii/S0022309319305885}{{Glass
  engineering to enhance Si solar cells: A case study of Pr3+-Yb3+ codoped
  tellurite-tungstate as spectral converter}}, Journal of Non-Crystalline
  Solids 526~(September) (2019) 119717.
\newblock \href {https://doi.org/10.1016/j.jnoncrysol.2019.119717}
  {\path{doi:10.1016/j.jnoncrysol.2019.119717}}.
\newline\urlprefix\url{10.1016/j.jnoncrysol.2019.119717
  https://linkinghub.elsevier.com/retrieve/pii/S0022309319305885}

\bibitem{Han2015a}
B.~Han, Y.~Yang, J.~Wu, J.~Wei, Z.~Li, Y.~Mai,
  \href{https://linkinghub.elsevier.com/retrieve/pii/S0272884215011669}{{Al2O3:Cr3+/tellurite
  glass composites: An efficient light converter for silicon solar cell}},
  Ceramics International 41~(9) (2015) 12267--12272.
\newblock \href {https://doi.org/10.1016/j.ceramint.2015.06.050}
  {\path{doi:10.1016/j.ceramint.2015.06.050}}.
\newline\urlprefix\url{https://linkinghub.elsevier.com/retrieve/pii/S0272884215011669}

\bibitem{Jiang2016}
X.~Zhou, J.~Shen, Y.~Wang, Z.~Feng, R.~Wang, L.~Li, S.~Jiang, X.~Luo,
  \href{http://doi.wiley.com/10.1111/jace.14133}{{An Efficient Dual-Mode Solar
  Spectral Modification for c-Si Solar Cells in Tm 3+ /Yb 3+ Codoped Tellurite
  Glasses}}, Journal of the American Ceramic Society 99~(7) (2016) 2300--2305.
\newblock \href {https://doi.org/10.1111/jace.14133}
  {\path{doi:10.1111/jace.14133}}.
\newline\urlprefix\url{http://doi.wiley.com/10.1111/jace.14133}

\bibitem{Garcia2019}
J.~A. Garcia, L.~Bontempo, L.~A. Gomez-Malagon, L.~R. Kassab,
  \href{https://doi.org/10.1016/j.optmat.2018.11.028}{{Efficiency boost in
  Si-based solar cells using tellurite glass cover layer doped with Eu3+ and
  silver nanoparticles}}, Optical Materials 88~(November 2018) (2019) 155--160.
\newblock \href {https://doi.org/10.1016/j.optmat.2018.11.028}
  {\path{doi:10.1016/j.optmat.2018.11.028}}.
\newline\urlprefix\url{https://doi.org/10.1016/j.optmat.2018.11.028}

\bibitem{Taniguchi2020}
M.~M. Taniguchi, E.~da~Silva, M.~A.~T. da~Silva, L.~S. Herculano, R.~F. Muniz,
  M.~Sandrini, M.~P. Belan{\c{c}}on,
  \href{https://doi.org/10.1016/j.jnoncrysol.2020.120307
  https://linkinghub.elsevier.com/retrieve/pii/S0022309320304191}{{The role of
  Ce3+/Ce4+ in the spectroscopic properties of cerium oxide doped
  zinc-tellurite glasses prepared under air}}, Journal of Non-Crystalline
  Solids 547~(July) (2020) 120307.
\newblock \href {https://doi.org/10.1016/j.jnoncrysol.2020.120307}
  {\path{doi:10.1016/j.jnoncrysol.2020.120307}}.
\newline\urlprefix\url{https://doi.org/10.1016/j.jnoncrysol.2020.120307
  https://linkinghub.elsevier.com/retrieve/pii/S0022309320304191}

\bibitem{Belancon2021}
M.~Belan{\c{c}}on, M.~Sandrini, H.~Muniz, L.~Herculano, G.~Lukasievicz,
  E.~Savi, O.~Capeloto, L.~Malacarne, N.~Astrath, M.~Baesso, G.~Schiavon,
  A.~{Silva Junior}, J.~Marconi,
  \href{https://doi.org/10.1016/j.jpcs.2021.110396
  https://linkinghub.elsevier.com/retrieve/pii/S0022369721004625}{{Float,
  borosilicate and tellurites as cover glasses in Si photovoltaics: Optical
  properties and performances under sunlight}}, Journal of Physics and
  Chemistry of Solids 161~(March 2021) (2022) 110396.
\newblock \href {https://doi.org/10.1016/j.jpcs.2021.110396}
  {\path{doi:10.1016/j.jpcs.2021.110396}}.
\newline\urlprefix\url{https://doi.org/10.1016/j.jpcs.2021.110396
  https://linkinghub.elsevier.com/retrieve/pii/S0022369721004625}

\bibitem{Bosio2020}
A.~Bosio, S.~Pasini, N.~Romeo,
  \href{https://www.mdpi.com/2079-6412/10/4/344}{{The History of Photovoltaics
  with Emphasis on CdTe Solar Cells and Modules}}, Coatings 10~(4) (2020) 344.
\newblock \href {https://doi.org/10.3390/coatings10040344}
  {\path{doi:10.3390/coatings10040344}}.
\newline\urlprefix\url{https://www.mdpi.com/2079-6412/10/4/344}

\bibitem{Graedel2011}
T.~Graedel,
  \href{http://www.annualreviews.org/doi/10.1146/annurev-matsci-062910-095759}{{On
  the Future Availability of the Energy Metals}}, Annual Review of Materials
  Research 41~(1) (2011) 323--335.
\newblock \href {https://doi.org/10.1146/annurev-matsci-062910-095759}
  {\path{doi:10.1146/annurev-matsci-062910-095759}}.
\newline\urlprefix\url{http://www.annualreviews.org/doi/10.1146/annurev-matsci-062910-095759}

\bibitem{Graedel2015a}
T.~E. Graedel, E.~M. Harper, N.~T. Nassar, P.~Nuss, B.~K. Reck, B.~L. Turner,
  {Criticality of metals and metalloids}, Proceedings of the National Academy
  of Sciences of the United States of America 112~(14) (2015) 4257--4262.
\newblock \href {https://doi.org/10.1073/pnas.1500415112}
  {\path{doi:10.1073/pnas.1500415112}}.

\bibitem{Zhang2017c}
K.~Zhang, A.~N. Kleit, A.~Nieto,
  \href{http://dx.doi.org/10.1016/j.rser.2016.12.127}{{An economics strategy
  for criticality – Application to rare earth element Yttrium in new lighting
  technology and its sustainable availability}}, Renewable and Sustainable
  Energy Reviews 77~(March 2016) (2017) 899--915.
\newblock \href {https://doi.org/10.1016/j.rser.2016.12.127}
  {\path{doi:10.1016/j.rser.2016.12.127}}.
\newline\urlprefix\url{http://dx.doi.org/10.1016/j.rser.2016.12.127}

\bibitem{rosales-sosa2016}
G.~A. Rosales-Sosa, A.~Masuno, Y.~Higo, H.~Inoue,
  \href{https://www.nature.com/articles/srep23620}{{Crack-resistant
  Al2O3–SiO2 glasses}}, Scientific Reports 6~(1) (2016) 23620.
\newblock \href {https://doi.org/10.1038/srep23620}
  {\path{doi:10.1038/srep23620}}.
\newline\urlprefix\url{https://www.nature.com/articles/srep23620}

\bibitem{DeSousa2003b}
D.~F. {De Sousa}, L.~A. Nunes, J.~H. Rohling, M.~L. Baesso, {Laser emission at
  1077 nm in Nd3+-doped calcium aluminosilicate glass}, Applied Physics B:
  Lasers and Optics 77~(1) (2003) 59--63.
\newblock \href {https://doi.org/10.1007/s00340-003-1247-y}
  {\path{doi:10.1007/s00340-003-1247-y}}.

\bibitem{Hess1996}
K.~U. Hess, D.~B. Dingwell, E.~R{\"{o}}ssler, {Parametrization of
  viscosity-temperature relations of aluminosilicate melts}, Chemical Geology
  128~(1-4) (1996) 155--163.
\newblock \href {https://doi.org/10.1016/0009-2541(95)00170-0}
  {\path{doi:10.1016/0009-2541(95)00170-0}}.

\bibitem{Hara2014}
K.~Hara, H.~Ichinose, T.~N. Murakami, A.~Masuda, {Crystalline Si photovoltaic
  modules based on TiO2-coated cover glass against potential-induced
  degradation}, RSC Advances 4~(83) (2014) 44291--44295.
\newblock \href {https://doi.org/10.1039/c4ra06791f}
  {\path{doi:10.1039/c4ra06791f}}.

\bibitem{Giolando2016}
D.~M. Giolando,
  \href{http://dx.doi.org/10.1016/j.solener.2015.11.024}{{Transparent
  self-cleaning coating applicable to solar energy consisting of nano-crystals
  of titanium dioxide in fluorine doped tin dioxide}}, Solar Energy 124 (2016)
  76--81.
\newblock \href {https://doi.org/10.1016/j.solener.2015.11.024}
  {\path{doi:10.1016/j.solener.2015.11.024}}.
\newline\urlprefix\url{http://dx.doi.org/10.1016/j.solener.2015.11.024}

\bibitem{Pratiwi2020}
N.~Pratiwi, Zulhadjri, S.~Arief, D.~V. Wellia, {A Facile Preparation of
  Transparent Ultrahydrophobic Glass via TiO2/Octadecyltrichlorosilane (ODTS)
  Coatings for Self-Cleaning Material}, ChemistrySelect 5~(4) (2020)
  1450--1454.
\newblock \href {https://doi.org/10.1002/slct.201904153}
  {\path{doi:10.1002/slct.201904153}}.

\bibitem{Khan2020}
S.~B. Khan, Z.~Zhang, S.~L. Lee, {Single component: Bilayer TiO2 as a durable
  antireflective coating}, Journal of Alloys and Compounds 834 (2020).
\newblock \href {https://doi.org/10.1016/j.jallcom.2020.155137}
  {\path{doi:10.1016/j.jallcom.2020.155137}}.

\bibitem{Saito2002}
K.~Saito, A.~J. Ikushima, {Effects of fluorine on structure, structural
  relaxation, and absorption edge in silica glass}, Journal of Applied Physics
  91~(8) (2002) 4886--4890.
\newblock \href {https://doi.org/10.1063/1.1459102}
  {\path{doi:10.1063/1.1459102}}.

\bibitem{Mukherjee2013}
D.~P. Mukherjee, S.~K. Das, {SiO 2-Al 2O 3-CaO glass-ceramics: Effects of CaF 2
  on crystallization, microstructure and properties}, Ceramics International
  39~(1) (2013) 571--578.
\newblock \href {https://doi.org/10.1016/j.ceramint.2012.06.066}
  {\path{doi:10.1016/j.ceramint.2012.06.066}}.

\bibitem{Pei2020}
F.~Pei, G.~Zhu, P.~Li, H.~Guo, P.~Yang, {Effects of CaF2 on the sintering and
  crystallisation of CaO–MgO–Al2O3–SiO2 glass-ceramics}, Ceramics
  International 46~(11) (2020) 17825--17835.
\newblock \href {https://doi.org/10.1016/j.ceramint.2020.04.089}
  {\path{doi:10.1016/j.ceramint.2020.04.089}}.

\bibitem{Muniz2021b}
R.~Muniz, V.~Soares, V.~Zanuto, M.~Melo, M.~Sandrini, M.~Belan{\c{c}}on,
  A.~Medina, M.~Baesso, \href{https://doi.org/10.1016/j.jnoncrysol.2021.121169
  https://linkinghub.elsevier.com/retrieve/pii/S0022309321005329}{{Color
  tunability and synergistic effect of PiG materials based on YAG:Ce3+ phosphor
  in SCS:Eu3+ glass}}, Journal of Non-Crystalline Solids 574~(September) (2021)
  121169.
\newblock \href {https://doi.org/10.1016/j.jnoncrysol.2021.121169}
  {\path{doi:10.1016/j.jnoncrysol.2021.121169}}.
\newline\urlprefix\url{https://doi.org/10.1016/j.jnoncrysol.2021.121169
  https://linkinghub.elsevier.com/retrieve/pii/S0022309321005329}

\bibitem{Kikuchi2022}
R.~Kikuchi, T.~Sato, N.~Fujii, M.~Shimbashi, C.~A. Arcilla,
  \href{https://doi.org/10.1038/s41598-022-20482-3}{{Natural glass alteration
  under a hyperalkaline condition for about 4000 years}}, Scientific Reports
  (2022) 1--10\href {https://doi.org/10.1038/s41598-022-20482-3}
  {\path{doi:10.1038/s41598-022-20482-3}}.
\newline\urlprefix\url{https://doi.org/10.1038/s41598-022-20482-3}

\bibitem{Guo2022}
Q.~Guo, T.~Feng, M.~J. Lance, K.~A. Unocic, S.~T. Pantelides, E.~Lara-Curzio,
  \href{https://doi.org/10.1016/j.jeurceramsoc.2021.11.013
  https://linkinghub.elsevier.com/retrieve/pii/S0955221921008116}{{Evolution of
  the structure and chemical composition of the interface between
  multi-component silicate glasses and yttria-stabilized zirconia after 40,000
  h exposure in air at 800 °C}}, Journal of the European Ceramic Society
  42~(4) (2022) 1576--1584.
\newblock \href {https://doi.org/10.1016/j.jeurceramsoc.2021.11.013}
  {\path{doi:10.1016/j.jeurceramsoc.2021.11.013}}.
\newline\urlprefix\url{https://doi.org/10.1016/j.jeurceramsoc.2021.11.013
  https://linkinghub.elsevier.com/retrieve/pii/S0955221921008116}

\bibitem{Gin2021}
S.~Gin, J.~M. Delaye, F.~Angeli, S.~Schuller,
  \href{http://dx.doi.org/10.1038/s41529-021-00190-5}{{Aqueous alteration of
  silicate glass: state of knowledge and perspectives}}, npj Materials
  Degradation 5~(1) (2021).
\newblock \href {https://doi.org/10.1038/s41529-021-00190-5}
  {\path{doi:10.1038/s41529-021-00190-5}}.
\newline\urlprefix\url{http://dx.doi.org/10.1038/s41529-021-00190-5}

\bibitem{Geisler2019}
T.~Geisler, L.~Dohmen, C.~Lenting, M.~B.~K. Fritzsche,
  \href{http://dx.doi.org/10.1038/s41563-019-0293-8
  https://www.nature.com/articles/s41563-019-0293-8}{{Real-time in situ
  observations of reaction and transport phenomena during silicate glass
  corrosion by fluid-cell Raman spectroscopy}}, Nature Materials 18~(4) (2019)
  342--348.
\newblock \href {https://doi.org/10.1038/s41563-019-0293-8}
  {\path{doi:10.1038/s41563-019-0293-8}}.
\newline\urlprefix\url{http://dx.doi.org/10.1038/s41563-019-0293-8
  https://www.nature.com/articles/s41563-019-0293-8}

\bibitem{Du2019a}
T.~Du, H.~Li, \href{http://dx.doi.org/10.1038/s41529-019-0070-9}{{Atomistic
  origin of the passivation effect in hydrated silicate glasses}}, npj
  Materials Degradation~(November 2018) (2019).
\newblock \href {https://doi.org/10.1038/s41529-019-0070-9}
  {\path{doi:10.1038/s41529-019-0070-9}}.
\newline\urlprefix\url{http://dx.doi.org/10.1038/s41529-019-0070-9}

\bibitem{Wang2018a}
Y.~Wang, C.~F. Jove-colon, C.~Lenting, J.~Icenhower, K.~L. Kuhlman,
  \href{http://dx.doi.org/10.1038/s41529-018-0047-0}{{Morphological instability
  of aqueous dissolution of silicate glasses and minerals}}, npj Materials
  Degradation~(August) (2018).
\newblock \href {https://doi.org/10.1038/s41529-018-0047-0}
  {\path{doi:10.1038/s41529-018-0047-0}}.
\newline\urlprefix\url{http://dx.doi.org/10.1038/s41529-018-0047-0}

\bibitem{Guiheneuf2017}
V.~Guiheneuf, F.~Delaleux, O.~Riou, P.~O. Logerais, J.~F. Durastanti,
  \href{http://dx.doi.org/10.1080/1478422X.2016.1234803}{{Investigation of damp
  heat effects on glass properties for photovoltaic applications}}, Corrosion
  Engineering Science and Technology 52~(3) (2017) 170--177.
\newblock \href {https://doi.org/10.1080/1478422X.2016.1234803}
  {\path{doi:10.1080/1478422X.2016.1234803}}.
\newline\urlprefix\url{http://dx.doi.org/10.1080/1478422X.2016.1234803}

\bibitem{Seshadri2022}
M.~Seshadri, I.~T. Santos, M.~J.~V. Bell, J.~Lapointe, Y.~Messaddeq, V.~Anjos,
  \href{https://doi.org/10.1038/s41598-022-23808-3}{{Near-infrared quantum
  cutting luminescence in Pr3+/Yb3+ doped lead bismuth borate glass}},
  Scientific Reports 12~(1) (2022) 1--8.
\newblock \href {https://doi.org/10.1038/s41598-022-23808-3}
  {\path{doi:10.1038/s41598-022-23808-3}}.
\newline\urlprefix\url{https://doi.org/10.1038/s41598-022-23808-3}

\bibitem{Romero-Romo2021a}
W.~Romero-Romo, S.~Carmona-T{\'{e}}llez, R.~Lozada-Morales, O.~Soriano-Romero,
  U.~Caldi{\~{n}}o, M.~E. {\'{A}}lvarez-Ramos, M.~E. Zayas, A.~N. Meza-Rocha,
  {Down-shifting and down-conversion emission properties of novel CdO–P2O5
  invert glasses activated with Pr3+ and Pr3+/Yb3+ for photonic applications},
  Optical Materials 116~(February) (2021).
\newblock \href {https://doi.org/10.1016/j.optmat.2021.111009}
  {\path{doi:10.1016/j.optmat.2021.111009}}.

\bibitem{Aouaini2022}
F.~Aouaini, A.~Maaoui, N.~B.~H. Mohamed, M.~M. Alanazi, L.~A. {El Maati},
  \href{https://doi.org/10.1016/j.jallcom.2021.162506}{{Visible to infrared
  down conversion of Er3+ doped tellurite glass for luminescent solar
  converters}}, Journal of Alloys and Compounds 894 (2022) 162506.
\newblock \href {https://doi.org/10.1016/j.jallcom.2021.162506}
  {\path{doi:10.1016/j.jallcom.2021.162506}}.
\newline\urlprefix\url{https://doi.org/10.1016/j.jallcom.2021.162506}

\bibitem{Kaniyarakkal2023}
S.~Kaniyarakkal, K.~Culala, R.~Dagupati,
  \href{https://doi.org/10.1016/j.bsecv.2023.01.002}{{Down conversion and
  efficient NIR to visible up-conversion emission analysis in Ho 3 + / Yb 3 +
  co-doped tellurite glasses}}, Bolet{\'{i}}n de la Sociedad Espa{\~{n}}ola de
  Cer{\'{a}}mica y Vidrio (2023) 1--8\href
  {https://doi.org/10.1016/j.bsecv.2023.01.002}
  {\path{doi:10.1016/j.bsecv.2023.01.002}}.
\newline\urlprefix\url{https://doi.org/10.1016/j.bsecv.2023.01.002}

\bibitem{Bouzidi2022a}
M.~Bouzidi, A.~Maaoui, N.~Chaaben, A.~S. Alshammari, Z.~R. Khan, M.~Mohamed,
  \href{https://doi.org/10.1016/j.jnoncrysol.2022.121837}{{Downconversion
  mechanism in Er3+/Yb3+ codoped fluorotellurite glasses to enhance the
  efficiency of c-Si PV cells}}, Journal of Non-Crystalline Solids 595~(July)
  (2022) 121837.
\newblock \href {https://doi.org/10.1016/j.jnoncrysol.2022.121837}
  {\path{doi:10.1016/j.jnoncrysol.2022.121837}}.
\newline\urlprefix\url{https://doi.org/10.1016/j.jnoncrysol.2022.121837}

\bibitem{Singh2023}
H.~Singh, T.~Singh, D.~Singh, V.~Bhatia, D.~Kumar, S.~P. Singh,
  \href{https://doi.org/10.1016/j.optmat.2023.113586}{{Up-conversion and
  downconversion studies of Nd3+ doped borophosphate glasses}}, Optical
  Materials 137~(February) (2023) 113586.
\newblock \href {https://doi.org/10.1016/j.optmat.2023.113586}
  {\path{doi:10.1016/j.optmat.2023.113586}}.
\newline\urlprefix\url{https://doi.org/10.1016/j.optmat.2023.113586}

\bibitem{Chen2018}
Y.~Chen, G.~Chen, X.~Liu, J.~Xu, T.~Yang, C.~Yuan, C.~Zhou,
  \href{https://doi.org/10.1016/j.jnoncrysol.2018.01.027}{{Down-conversion
  luminescence and optical thermometric performance of Tb3+/Eu3+ doped
  phosphate glass}}, Journal of Non-Crystalline Solids 484~(January) (2018)
  111--117.
\newblock \href {https://doi.org/10.1016/j.jnoncrysol.2018.01.027}
  {\path{doi:10.1016/j.jnoncrysol.2018.01.027}}.
\newline\urlprefix\url{https://doi.org/10.1016/j.jnoncrysol.2018.01.027}

\bibitem{Benrejeb2022}
H.~Benrejeb, K.~Soler-Carracedo, S.~Hraiech, I.~R. Martin,
  \href{https://doi.org/10.1016/j.optmat.2022.112604}{{Analysis of down
  conversion and back-transfer processes in Pr3+-Yb3+ co-doped phosphate
  glasses}}, Optical Materials 131~(June) (2022) 112604.
\newblock \href {https://doi.org/10.1016/j.optmat.2022.112604}
  {\path{doi:10.1016/j.optmat.2022.112604}}.
\newline\urlprefix\url{https://doi.org/10.1016/j.optmat.2022.112604}

\bibitem{Zhang2022}
T.~Zhang, D.~Zhang, P.~A. Wang, C.~Cui, {Preparation of Dy3+/Tm3+Co-doped
  Phosphate Glasses by Melt Method and its Luminescence Properties}, Journal of
  Physics: Conference Series 2226~(1) (2022).
\newblock \href {https://doi.org/10.1088/1742-6596/2226/1/012004}
  {\path{doi:10.1088/1742-6596/2226/1/012004}}.

\bibitem{Shi2020}
D.~Y. Shi, S.~bao Lin, X.~xia Zhao, A.~ling Feng, Q.~Xu,
  \href{https://doi.org/10.1016/j.rinp.2020.103411}{{Near-infrared quantum
  cutting in Tm3+/Yb3+-doped phosphate glasses}}, Results in Physics 19 (2020)
  103411.
\newblock \href {https://doi.org/10.1016/j.rinp.2020.103411}
  {\path{doi:10.1016/j.rinp.2020.103411}}.
\newline\urlprefix\url{https://doi.org/10.1016/j.rinp.2020.103411}

\bibitem{Markvart2008}
T.~Markvart, {Solar cell as a heat engine: Energy-entropy analysis of
  photovoltaic conversion}, Physica Status Solidi (A) Applications and
  Materials Science 205~(12) (2008) 2752--2756.
\newblock \href {https://doi.org/10.1002/pssa.200880460}
  {\path{doi:10.1002/pssa.200880460}}.

\bibitem{Markvart2016}
T.~Markvart, {From steam engine to solar cells: can thermodynamics guide the
  development of future generations of photovoltaics?}, Wiley Interdisciplinary
  Reviews: Energy and Environment 5~(5) (2016) 543--569.
\newblock \href {https://doi.org/10.1002/wene.204}
  {\path{doi:10.1002/wene.204}}.

\bibitem{Chu2012}
S.~Chu, A.~Majumdar,
  \href{http://www.nature.com/doifinder/10.1038/nature11475}{{Opportunities and
  challenges for a sustainable energy future}}, Nature 488~(7411) (2012)
  294--303.
\newblock \href {https://doi.org/10.1038/nature11475}
  {\path{doi:10.1038/nature11475}}.
\newline\urlprefix\url{http://www.nature.com/doifinder/10.1038/nature11475}

\bibitem{Xu2015}
Y.~Xu, T.~Gong, J.~N. Munday, \href{http://dx.doi.org/10.1038/srep13536}{{The
  generalized Shockley-Queisser limit for nanostructured solar cells}}, Nature
  Publishing Group (2015) 1--9\href {https://doi.org/10.1038/srep13536}
  {\path{doi:10.1038/srep13536}}.
\newline\urlprefix\url{http://dx.doi.org/10.1038/srep13536}

\bibitem{Kim2020a}
M.~S. Kim, J.~H. Lee, M.~K. Kwak,
  \href{https://doi.org/10.1007/s12541-020-00337-5}{{Review: Surface Texturing
  Methods for Solar Cell Efficiency Enhancement}}, International Journal of
  Precision Engineering and Manufacturing 21~(7) (2020) 1389--1398.
\newblock \href {https://doi.org/10.1007/s12541-020-00337-5}
  {\path{doi:10.1007/s12541-020-00337-5}}.
\newline\urlprefix\url{https://doi.org/10.1007/s12541-020-00337-5}

\bibitem{Zhou2021}
Z.~Zhou, Y.~Jiang, N.~Ekins-Daukes, M.~Keevers, M.~A. Green, {Optical and
  Thermal Emission Benefits of Differently Textured Glass for Photovoltaic
  Modules}, IEEE Journal of Photovoltaics 11~(1) (2021) 131--137.
\newblock \href {https://doi.org/10.1109/JPHOTOV.2020.3033390}
  {\path{doi:10.1109/JPHOTOV.2020.3033390}}.

\bibitem{Buskens2016}
P.~Buskens, M.~Burghoorn, M.~C.~D. Mourad, Z.~Vroon, {Antireflective Coatings
  for Glass and Transparent Polymers}, Langmuir 32~(27) (2016) 6781--6793.
\newblock \href {https://doi.org/10.1021/acs.langmuir.6b00428}
  {\path{doi:10.1021/acs.langmuir.6b00428}}.

\bibitem{Lobmann2018}
P.~L{\"{o}}bmann, {Sol-Gel Processing of MgF2 Antireflective Coatings},
  Nanomaterials 8~(5) (2018) 295.
\newblock \href {https://doi.org/10.3390/nano8050295}
  {\path{doi:10.3390/nano8050295}}.

\bibitem{Morales2018}
F.~Wiesinger, G.~S. Vicente, A.~Fern{\'{a}}ndez-Garc{\'{i}}a, F.~Sutter,
  {\'{A}}.~Morales, R.~Pitz-Paal,
  \href{https://linkinghub.elsevier.com/retrieve/pii/S0927024818300692}{{Sandstorm
  erosion testing of anti-reflective glass coatings for solar energy
  applications}}, Solar Energy Materials and Solar Cells 179~(September 2017)
  (2018) 10--16.
\newblock \href {https://doi.org/10.1016/j.solmat.2018.02.018}
  {\path{doi:10.1016/j.solmat.2018.02.018}}.
\newline\urlprefix\url{https://linkinghub.elsevier.com/retrieve/pii/S0927024818300692}

\bibitem{Womack2019}
G.~Womack, K.~Isbilir, F.~Lisco, G.~Durand, A.~Taylor, J.~M. Walls,
  \href{https://doi.org/10.1016/j.surfcoat.2018.11.030}{{The performance and
  durability of single-layer sol-gel anti-reflection coatings applied to solar
  module cover glass}}, Surface and Coatings Technology 358~(November 2018)
  (2019) 76--83.
\newblock \href {https://doi.org/10.1016/j.surfcoat.2018.11.030}
  {\path{doi:10.1016/j.surfcoat.2018.11.030}}.
\newline\urlprefix\url{https://doi.org/10.1016/j.surfcoat.2018.11.030}

\bibitem{Maghami2016a}
M.~R. Maghami, H.~Hizam, C.~Gomes, M.~A. Radzi, M.~I. Rezadad, S.~Hajighorbani,
  \href{http://dx.doi.org/10.1016/j.rser.2016.01.044}{{Power loss due to
  soiling on solar panel: A review}}, Renewable and Sustainable Energy Reviews
  59 (2016) 1307--1316.
\newblock \href {https://doi.org/10.1016/j.rser.2016.01.044}
  {\path{doi:10.1016/j.rser.2016.01.044}}.
\newline\urlprefix\url{http://dx.doi.org/10.1016/j.rser.2016.01.044}

\bibitem{Syafiq2018}
A.~Syafiq, A.~K. Pandey, N.~N. Adzman, N.~A. Rahim, {Advances in approaches and
  methods for self-cleaning of solar photovoltaic panels}, Solar Energy
  162~(May 2017) (2018) 597--619.
\newblock \href {https://doi.org/10.1016/j.solener.2017.12.023}
  {\path{doi:10.1016/j.solener.2017.12.023}}.

\bibitem{Lu2020}
H.~Lu, R.~Cai, L.~Z. Zhang, L.~Lu, L.~Zhang,
  \href{https://doi.org/10.1016/j.solener.2020.06.012}{{Experimental
  investigation on deposition reduction of different types of dust on solar PV
  cells by self-cleaning coatings}}, Solar Energy 206~(May) (2020) 365--373.
\newblock \href {https://doi.org/10.1016/j.solener.2020.06.012}
  {\path{doi:10.1016/j.solener.2020.06.012}}.
\newline\urlprefix\url{https://doi.org/10.1016/j.solener.2020.06.012}

\bibitem{cheng2021}
H.~Cheng, F.~Wang, J.~Ou, W.~Li, R.~Xue, {Solar reflective coatings with
  luminescence and self-cleaning function}, Surfaces and Interfaces 26 (oct
  2021).
\newblock \href {https://doi.org/10.1016/j.surfin.2021.101325}
  {\path{doi:10.1016/j.surfin.2021.101325}}.

\bibitem{biswas2022}
D.~Biswas, N.~Chundi, S.~R. Atchuta, K.~K. {Phani Kumar}, M.~{Shiva Prasad},
  S.~Sakthivel, {Fabrication of omnidirectional broadband dual-functional
  coating with high optical and self-cleaning properties for photovoltaic
  application}, Solar Energy 246 (2022) 36--44.
\newblock \href {https://doi.org/10.1016/j.solener.2022.09.038}
  {\path{doi:10.1016/j.solener.2022.09.038}}.

\bibitem{Phys2022}
A.~C. Bento, N.~Cella, S.~M. Lima, L.~A.~O. Nunes, L.~H.~C. Andrade, J.~R.
  Silva, V.~S. Zanuto, N.~G.~C. Astrath, T.~Catunda, A.~N. Medina, J.~H.
  Rohling, R.~F. Muniz, J.~W. Berrar, L.~C. Malacarne, W.~R. Weinand, F.~Sato,
  M.~P. Belancon, G.~J. Schiavon, J.~Shen, L.~C.~M. Miranda, H.~Vargas, M.~L.
  Baesso, \href{https://aip.scitation.org/doi/10.1063/5.0088211}{{Photoacoustic
  and photothermal and the photovoltaic efficiency of solar cells: A
  tutorial}}, Journal of Applied Physics 131~(14) (2022) 141101.
\newblock \href {https://doi.org/10.1063/5.0088211}
  {\path{doi:10.1063/5.0088211}}.
\newline\urlprefix\url{https://aip.scitation.org/doi/10.1063/5.0088211}

\bibitem{Huang2013}
X.~Huang, S.~Han, W.~Huang, X.~Liu,
  \href{http://xlink.rsc.org/?DOI=C2CS35288E}{{Enhancing solar cell efficiency:
  the search for luminescent materials as spectral converters}}, Chemical
  Society Reviews 42~(1) (2013) 173--201.
\newblock \href {https://doi.org/10.1039/c2cs35288e}
  {\path{doi:10.1039/c2cs35288e}}.
\newline\urlprefix\url{http://xlink.rsc.org/?DOI=C2CS35288E}

\bibitem{tawalare2021}
P.~K. Tawalare,
  \href{https://aip.scitation.org/doi/10.1063/5.0064202}{{Optimizing
  photovoltaic conversion of solar energy}}, AIP Advances 11~(10) (2021)
  100701.
\newblock \href {https://doi.org/10.1063/5.0064202}
  {\path{doi:10.1063/5.0064202}}.
\newline\urlprefix\url{https://aip.scitation.org/doi/10.1063/5.0064202}

\bibitem{Ghazy2021}
A.~Ghazy, M.~Safdar, M.~Lastusaari, H.~Savin, M.~Karppinen,
  \href{https://doi.org/10.1016/j.solmat.2021.111234}{{Advances in upconversion
  enhanced solar cell performance}}, Solar Energy Materials and Solar Cells
  230~(April) (2021) 111234.
\newblock \href {https://doi.org/10.1016/j.solmat.2021.111234}
  {\path{doi:10.1016/j.solmat.2021.111234}}.
\newline\urlprefix\url{https://doi.org/10.1016/j.solmat.2021.111234}

\bibitem{Khare2020}
A.~Khare, \href{https://doi.org/10.1016/j.jallcom.2019.153214}{{A critical
  review on the efficiency improvement of upconversion assisted solar cells}},
  Journal of Alloys and Compounds 821 (2020) 153214.
\newblock \href {https://doi.org/10.1016/j.jallcom.2019.153214}
  {\path{doi:10.1016/j.jallcom.2019.153214}}.
\newline\urlprefix\url{https://doi.org/10.1016/j.jallcom.2019.153214}

\bibitem{Lin2016c}
C.~C. Lin, P.~J. Krommenhoek, S.~S. Watson, X.~Gu,
  \href{http://dx.doi.org/10.1016/j.solmat.2015.09.021}{{Depth profiling of
  degradation of multilayer photovoltaic backsheets after accelerated
  laboratory weathering: Cross-sectional Raman imaging}}, Solar Energy
  Materials and Solar Cells 144 (2016) 289--299.
\newblock \href {https://doi.org/10.1016/j.solmat.2015.09.021}
  {\path{doi:10.1016/j.solmat.2015.09.021}}.
\newline\urlprefix\url{http://dx.doi.org/10.1016/j.solmat.2015.09.021}

\bibitem{Schlothauer2012}
J.~Schlothauer, S.~Jungwirth, M.~K{\"{o}}hl, B.~R{\"{o}}der, {Degradation of
  the encapsulant polymer in outdoor weathered photovoltaic modules: Spatially
  resolved inspection of EVA ageing by fluorescence and correlation to
  electroluminescence}, Solar Energy Materials and Solar Cells 102 (2012)
  75--85.
\newblock \href {https://doi.org/10.1016/j.solmat.2012.03.022}
  {\path{doi:10.1016/j.solmat.2012.03.022}}.

\bibitem{Katayama2019}
N.~Katayama, S.~Osawa, S.~Matsumoto, T.~Nakano, M.~Sugiyama,
  \href{https://doi.org/10.1016/j.solmat.2019.01.040}{{Degradation and fault
  diagnosis of photovoltaic cells using impedance spectroscopy}}, Solar Energy
  Materials and Solar Cells 194~(September 2018) (2019) 130--136.
\newblock \href {https://doi.org/10.1016/j.solmat.2019.01.040}
  {\path{doi:10.1016/j.solmat.2019.01.040}}.
\newline\urlprefix\url{https://doi.org/10.1016/j.solmat.2019.01.040}

\bibitem{Zhang2017b}
K.~Zhang, L.~Hao, M.~Du, J.~Mi, J.-N. Wang, J.-p. Meng,
  \href{http://dx.doi.org/10.1016/j.rser.2016.09.083
  https://linkinghub.elsevier.com/retrieve/pii/S1364032116305731}{{A review on
  thermal stability and high temperature induced ageing mechanisms of solar
  absorber coatings}}, Renewable and Sustainable Energy Reviews 67 (2017)
  1282--1299.
\newblock \href {https://doi.org/10.1016/j.rser.2016.09.083}
  {\path{doi:10.1016/j.rser.2016.09.083}}.
\newline\urlprefix\url{http://dx.doi.org/10.1016/j.rser.2016.09.083
  https://linkinghub.elsevier.com/retrieve/pii/S1364032116305731}

\bibitem{francisnara2022}
M.~P. Belan{\c{c}}on, M.~Sandrini, F.~Tonholi, L.~S. Herculano, G.~S. Dias,
  \href{https://linkinghub.elsevier.com/retrieve/pii/S1755008422000515}{{Towards
  long term sustainability of c-Si solar panels: The environmental benefits of
  glass sheet recovery}}, Renewable Energy Focus 42 (2022) 206--210.
\newblock \href {https://doi.org/10.1016/j.ref.2022.06.009}
  {\path{doi:10.1016/j.ref.2022.06.009}}.
\newline\urlprefix\url{https://linkinghub.elsevier.com/retrieve/pii/S1755008422000515}

\bibitem{Gopalakrishna2019}
H.~Gopalakrishna, P.~Arularasu, K.~Dolia, A.~Sinha, G.~Tamizhmani,
  {Characterization of Encapsulant Degradation in Accelerated UV Stressed
  Mini-Modules with UV-cut and UV-pass EVA}, Conference Record of the IEEE
  Photovoltaic Specialists Conference (2019) 1961--1964\href
  {https://doi.org/10.1109/PVSC40753.2019.8980897}
  {\path{doi:10.1109/PVSC40753.2019.8980897}}.

\bibitem{dechaoyu2022}
D.~Yu, T.~Yu, H.~Lin, S.~Zhuang, D.~Zhang,
  \href{https://onlinelibrary.wiley.com/doi/10.1002/adom.202200014}{{Recent
  Advances in Luminescent Downconversion: New Materials, Techniques, and
  Applications in Solar Cells}}, Advanced Optical Materials 10~(12) (2022)
  2200014.
\newblock \href {https://doi.org/10.1002/adom.202200014}
  {\path{doi:10.1002/adom.202200014}}.
\newline\urlprefix\url{https://onlinelibrary.wiley.com/doi/10.1002/adom.202200014}

\bibitem{satpute2022}
N.~S. Satpute, C.~M. Mehare, A.~Tiwari, H.~C. Swart, S.~J. Dhoble,
  \href{https://pubs.acs.org/doi/10.1021/acsaelm.2c00595}{{Synthesis and
  Luminescence Characterization of Downconversion and Downshifting Phosphor for
  Efficiency Enhancement of Solar Cells: Perspectives and Challenges}}, ACS
  Applied Electronic Materials 4~(7) (2022) 3354--3391.
\newblock \href {https://doi.org/10.1021/acsaelm.2c00595}
  {\path{doi:10.1021/acsaelm.2c00595}}.
\newline\urlprefix\url{https://pubs.acs.org/doi/10.1021/acsaelm.2c00595}

\bibitem{lupei2014}
A.~Lupei, V.~Lupei, C.~Gheorghe, S.~Hau, A.~Ikesue, {Multicenters in Ce3+
  visible emission of YAG ceramics}, Optical Materials 37~(C) (2014) 727--733.
\newblock \href {https://doi.org/10.1016/j.optmat.2014.09.001}
  {\path{doi:10.1016/j.optmat.2014.09.001}}.

\bibitem{Teng2020a}
L.~Teng, Y.~Jiang, W.~Zhang, R.~Wei, H.~Guo, {Highly transparent cerium doped
  glasses with full-band UV-shielding capacity}, Journal of the American
  Ceramic Society 103~(5) (2020) 3249--3256.
\newblock \href {https://doi.org/10.1111/jace.17020}
  {\path{doi:10.1111/jace.17020}}.

\bibitem{Annapurna2004}
K.~Annapurna, R.~N. Dwivedi, P.~Kundu, S.~Buddhudu, {Blue emission spectrum of
  Ce3+:ZnO-B2O 3-SiO2 optical glass}, Materials Letters 58~(5) (2004) 787--789.
\newblock \href {https://doi.org/10.1016/j.matlet.2003.07.012}
  {\path{doi:10.1016/j.matlet.2003.07.012}}.

\bibitem{pullaiah2022}
G.~Pullaiah, K.~{Venkata Rao}, B.~C. Jamalaiah, N.~Madhu, V.~Nutalapati,
  {Spectroscopic and luminescent properties of Ce3+ -doped TeO2-WO3-GeO2
  glasses}, Materials Science and Engineering B: Solid-State Materials for
  Advanced Technology 284 (oct 2022).
\newblock \href {https://doi.org/10.1016/j.mseb.2022.115879}
  {\path{doi:10.1016/j.mseb.2022.115879}}.

\bibitem{Andrade2009}
L.~H.~C. Andrade, S.~M. Lima, a.~Novatski, a.~Steimacher, J.~H. Rohling, a.~N.
  Medina, a.~C. Bento, M.~L. Baesso, Y.~Guyot, G.~Boulon, {A step forward
  toward smart white lighting: Combination of glass phosphor and light emitting
  diodes}, Applied Physics Letters 95~(8) (2009) 2007--2010.
\newblock \href {https://doi.org/10.1063/1.3186784}
  {\path{doi:10.1063/1.3186784}}.

\bibitem{taizheng2015}
Y.~Tai, G.~Zheng, H.~Wang, J.~Bai, {Near-infrared quantum cutting of Ce3+-Nd3+
  co-doped Y3Al5O12 crystal for crystalline silicon solar cells}, Journal of
  Photochemistry and Photobiology A: Chemistry 303-304 (2015) 80--85.
\newblock \href {https://doi.org/10.1016/j.jphotochem.2015.02.009}
  {\path{doi:10.1016/j.jphotochem.2015.02.009}}.

\bibitem{wangqiu2015}
Q.~Wang, J.~B. Qiu, Z.~G. Song, Z.~W. Yang, Z.~Y. Yin, D.~C. Zhou, {Optical
  properties of Ce3+-Nd3+ co-doped YAG nanoparticles for visual and
  near-infrared biological imaging}, Spectrochimica Acta - Part A: Molecular
  and Biomolecular Spectroscopy 149 (2015) 898--903.
\newblock \href {https://doi.org/10.1016/j.saa.2015.04.082}
  {\path{doi:10.1016/j.saa.2015.04.082}}.

\bibitem{zhoulei2022}
Z.~Zhou, W.~Lei, P.~Zhang, H.~Liang, Z.~Luo, A.~Lu, {Influence of CaF2 addition
  on structure and luminescence properties of the
  Na2O–CaO–SiO2–Al2O3–ZnO–P2O5 glass co-doped with Ce3+/Yb3+},
  Optical Materials 134 (dec 2022).
\newblock \href {https://doi.org/10.1016/j.optmat.2022.113171}
  {\path{doi:10.1016/j.optmat.2022.113171}}.

\bibitem{Reddappa2021}
R.~Reddappa, K.~Suresh, C.~K. Jayasankar,
  \href{https://doi.org/10.1016/j.optmat.2021.111700}{{Down conversion studies
  in Ce3+ and Yb3+ doped Ca2SiO4 phosphors from agricultural waste: Si based
  solar cell applications}}, Optical Materials 122~(PB) (2021) 111700.
\newblock \href {https://doi.org/10.1016/j.optmat.2021.111700}
  {\path{doi:10.1016/j.optmat.2021.111700}}.
\newline\urlprefix\url{https://doi.org/10.1016/j.optmat.2021.111700}

\bibitem{pathak2017}
A.~A. Pathak, R.~A. Talewar, C.~P. Joshi, S.~V. Moharil, {Sensitization of Yb3+
  emission in CaYAl3O7 host}, Optical Materials 64 (2017) 217--223.
\newblock \href {https://doi.org/10.1016/j.optmat.2016.12.018}
  {\path{doi:10.1016/j.optmat.2016.12.018}}.

\bibitem{sontakke2016}
A.~D. Sontakke, J.~Ueda, S.~Tanabe, {Effect of synthesis conditions on Ce3 +
  luminescence in borate glasses}, Journal of Non-Crystalline Solids 431 (2016)
  150--153.
\newblock \href {https://doi.org/10.1016/j.jnoncrysol.2015.04.005}
  {\path{doi:10.1016/j.jnoncrysol.2015.04.005}}.

\bibitem{ranasinghe2022}
K.~S. Ranasinghe, R.~Singh, D.~Leshchev, A.~Vasquez, E.~Stavitski, I.~Foster,
  {Synthesis of Nanoceria with Varied Ratios of Ce3+/Ce4+ Utilizing Soluble
  Borate Glass}, Nanomaterials 12~(14) (jul 2022).
\newblock \href {https://doi.org/10.3390/nano12142363}
  {\path{doi:10.3390/nano12142363}}.

\bibitem{kaewnuam2022}
E.~Kaewnuam, N.~Wantana, Y.~Ruangtaweep, M.~Cadatal-Raduban, K.~Yamanoi, H.~J.
  Kim, P.~Kidkhunthod, J.~Kaewkhao, {The influence of CeF3 on radiation
  hardness and luminescence properties of Gd2O3–B2O3 glass scintillator},
  Scientific Reports 12~(1) (dec 2022).
\newblock \href {https://doi.org/10.1038/s41598-022-14833-3}
  {\path{doi:10.1038/s41598-022-14833-3}}.

\bibitem{song2012}
P.~Song, C.~Jiang, {Modeling of downconverter based on Pr3+-Yb3+ codoped
  fluoride glasses to improve sc-Si solar cells efficiency}, AIP Advances 2~(4)
  (dec 2012).
\newblock \href {https://doi.org/10.1063/1.4766187}
  {\path{doi:10.1063/1.4766187}}.

\bibitem{zhaozhao2020}
J.~Zhao, X.~Zhao, Z.~Leng, M.~Han, {Efficient blue to near-infrared
  luminescence properties in Pr3+ - Yb3+ co-doped Li8Bi2(MoO4)7 phosphor},
  Optical Materials 108 (oct 2020).
\newblock \href {https://doi.org/10.1016/j.optmat.2020.110232}
  {\path{doi:10.1016/j.optmat.2020.110232}}.

\bibitem{zhangcui2016}
G.~Zhang, Q.~Cui, G.~Liu, {Efficient near-infrared quantum cutting and
  downshift in Ce3+-Pr3+ codoped SrLaGa3S6O suitable for solar spectral
  converter}, Optical Materials 53 (2016) 214--217.
\newblock \href {https://doi.org/10.1016/j.optmat.2016.01.042}
  {\path{doi:10.1016/j.optmat.2016.01.042}}.

\bibitem{Meejitpaisan2021}
P.~Meejitpaisan, R.~Doddoji, S.~Kothan, C.~K. Jayasankar, J.~Kaewkhao, {Visible
  to infrared emission from (Eu3+/Nd3+):B2O3 + AlF3 + NaF + CaF2 glasses for
  luminescent solar converters}, Optics and Laser Technology 141~(December
  2020) (2021).
\newblock \href {https://doi.org/10.1016/j.optlastec.2021.107170}
  {\path{doi:10.1016/j.optlastec.2021.107170}}.

\bibitem{Luo2019}
J.~Luo, S.~ichi Amma, L.~Chen, D.~Ngo, J.~C. Mauro, C.~G. Pantano, S.~H. Kim,
  \href{https://doi.org/10.1016/j.jnoncrysol.2019.01.012}{{Relative abundance
  of subsurface hydroxyl and molecular water species in silicate and
  aluminosilicate glasses}}, Journal of Non-Crystalline Solids 510~(January)
  (2019) 179--185.
\newblock \href {https://doi.org/10.1016/j.jnoncrysol.2019.01.012}
  {\path{doi:10.1016/j.jnoncrysol.2019.01.012}}.
\newline\urlprefix\url{https://doi.org/10.1016/j.jnoncrysol.2019.01.012}

\bibitem{dorenbos2003}
P.~Dorenbos,
  \href{https://linkinghub.elsevier.com/retrieve/pii/S0022231303000784}{{Energy
  of the first 4f7→4f65d transition of Eu2+ in inorganic compounds}}, Journal
  of Luminescence 104~(4) (2003) 239--260.
\newblock \href {https://doi.org/10.1016/S0022-2313(03)00078-4}
  {\path{doi:10.1016/S0022-2313(03)00078-4}}.
\newline\urlprefix\url{https://linkinghub.elsevier.com/retrieve/pii/S0022231303000784}

\bibitem{dorenbos2003a}
P.~Dorenbos,
  \href{https://iopscience.iop.org/article/10.1088/0953-8984/15/27/311}{{Relation
  between Eu 2 and Ce 3 f d-transition energies in inorganic compounds}},
  Journal of Physics: Condensed Matter 15~(27) (2003) 4797--4807.
\newblock \href {https://doi.org/10.1088/0953-8984/15/27/311}
  {\path{doi:10.1088/0953-8984/15/27/311}}.
\newline\urlprefix\url{https://iopscience.iop.org/article/10.1088/0953-8984/15/27/311}

\bibitem{chenwang2017}
Y.~Chen, J.~Wang, M.~Zhang, Q.~Zeng, {Light conversion material: LiBaPO4:Eu2+,
  Pr3+, suitable for solar cell}, RSC Advances 7~(34) (2017) 21221--21225.
\newblock \href {https://doi.org/10.1039/c7ra01834g}
  {\path{doi:10.1039/c7ra01834g}}.

\bibitem{wang2015}
C.~Wang, T.~Xuan, J.~Liu, H.~Li, Z.~Sun, {Long Afterglow SrAl2O4:Eu2+,Dy3+
  Phosphors as Luminescent Down-Shifting Layer for Crystalline Silicon Solar
  Cells}, International Journal of Applied Ceramic Technology 12~(4) (2015)
  722--727.
\newblock \href {https://doi.org/10.1111/ijac.12281}
  {\path{doi:10.1111/ijac.12281}}.

\bibitem{talewar2016}
R.~A. Talewar, C.~P. Joshi, S.~V. Moharil, {Near infrared emission and energy
  transfer in Eu2+ - Nd3+ co-doped Ca2BO3Cl}, Optical Materials 55 (2016)
  44--48.
\newblock \href {https://doi.org/10.1016/j.optmat.2016.03.007}
  {\path{doi:10.1016/j.optmat.2016.03.007}}.

\bibitem{luo2019b}
X.~Luo, J.~Y. Ahn, S.~H. Kim,
  \href{https://linkinghub.elsevier.com/retrieve/pii/S0038092X18312040}{{Aerosol
  synthesis and luminescent properties of CaAl2O4:Eu2+, Nd3+ down-conversion
  phosphor particles for enhanced light harvesting of dye-sensitized solar
  cells}}, Solar Energy 178 (2019) 173--180.
\newblock \href {https://doi.org/10.1016/j.solener.2018.12.029}
  {\path{doi:10.1016/j.solener.2018.12.029}}.
\newline\urlprefix\url{https://linkinghub.elsevier.com/retrieve/pii/S0038092X18312040}

\bibitem{zhoutenk2011}
J.~Zhou, Y.~Teng, X.~Liu, Z.~Ma, J.~Qiu, {Broadband spectral conversion of
  visible light to near-infrared emission via energy transfer from Ce3+ to
  Nd3+/Yb3+ in YAG}, Journal of Materials Research 26~(5) (2011) 689--692.
\newblock \href {https://doi.org/10.1557/jmr.2010.84}
  {\path{doi:10.1557/jmr.2010.84}}.

\bibitem{Romero-Romo2021}
W.~Romero-Romo, S.~Carmona-T{\'{e}}llez, R.~Lozada-Morales, O.~Soriano-Romero,
  U.~Caldi{\~{n}}o, M.~E. {\'{A}}lvarez-Ramos, M.~E. Zayas, A.~N. Meza-Rocha,
  {Down-shifting and down-conversion emission properties of novel CdO–P2O5
  invert glasses activated with Pr3+ and Pr3+/Yb3+ for photonic applications},
  Optical Materials 116~(March) (2021).
\newblock \href {https://doi.org/10.1016/j.optmat.2021.111009}
  {\path{doi:10.1016/j.optmat.2021.111009}}.

\bibitem{Zhou2016}
X.~Zhou, J.~Shen, Y.~Wang, Z.~Feng, R.~Wang, L.~Li, S.~Jiang, X.~Luo, {An
  Efficient Dual-Mode Solar Spectral Modification for c-Si Solar Cells in
  Tm3+/Yb3+ Codoped Tellurite Glasses}, Journal of the American Ceramic Society
  99~(7) (2016) 2300--2305.
\newblock \href {https://doi.org/10.1111/jace.14133}
  {\path{doi:10.1111/jace.14133}}.

\bibitem{Elleuch2015}
R.~Elleuch, R.~Salhi, J.-L. Deschanvres, R.~Maalej,
  \href{http://aip.scitation.org/doi/10.1063/1.4906976}{{Antireflective
  downconversion ZnO:Er3+, Yb3+thin film for Si solar cell applications}},
  Journal of Applied Physics 117~(5) (2015) 055301.
\newblock \href {https://doi.org/10.1063/1.4906976}
  {\path{doi:10.1063/1.4906976}}.
\newline\urlprefix\url{http://aip.scitation.org/doi/10.1063/1.4906976}

\bibitem{Bouzidi2022}
M.~Bouzidi, A.~Maaoui, N.~Chaaben, A.~S. Alshammari, Z.~R. Khan, M.~Mohamed,
  \href{https://doi.org/10.1016/j.jnoncrysol.2022.121837}{{Downconversion
  mechanism in Er3+/Yb3+ codoped fluorotellurite glasses to enhance the
  efficiency of c-Si PV cells}}, Journal of Non-Crystalline Solids 595~(July)
  (2022) 121837.
\newblock \href {https://doi.org/10.1016/j.jnoncrysol.2022.121837}
  {\path{doi:10.1016/j.jnoncrysol.2022.121837}}.
\newline\urlprefix\url{https://doi.org/10.1016/j.jnoncrysol.2022.121837}

\bibitem{Saad2022}
N.~Saad, M.~Ibrahim, K.~H. Sadok, M.~Haouari,
  \href{https://doi.org/10.1016/j.jnoncrysol.2022.121707}{{Fluoroborophosphate
  glasses doped with Cr3+, Nd3+ and Yb3+as efficient light converters for
  silicon based solar cells}}, Journal of Non-Crystalline Solids 591~(May)
  (2022) 121707.
\newblock \href {https://doi.org/10.1016/j.jnoncrysol.2022.121707}
  {\path{doi:10.1016/j.jnoncrysol.2022.121707}}.
\newline\urlprefix\url{https://doi.org/10.1016/j.jnoncrysol.2022.121707}

\bibitem{kempe2009}
M.~D. Kempe, T.~Moricone, M.~Kilkenny, {Effects of Cerium Removal from Glass on
  Photovoltaic Module Performance and Stability Preprint}~(September) (2009).

\bibitem{Dan2023}
H.~K. Dan, N.~D. Trung, D.~Zhou, J.~Qiu,
  \href{https://doi.org/10.1016/j.jnoncrysol.2022.122086}{{Influences of Mn2+
  ions, and Mn2+–Yb3+ dimer on the optical band gaps and bandwidth flatness
  of near-infrared emissions of Ho3+/Tm3+, Ho3+/Tm3+/Yb3+ co-doped calcium
  aluminosilicate glasses}}, Journal of Non-Crystalline Solids 603~(October
  2022) (2023) 122086.
\newblock \href {https://doi.org/10.1016/j.jnoncrysol.2022.122086}
  {\path{doi:10.1016/j.jnoncrysol.2022.122086}}.
\newline\urlprefix\url{https://doi.org/10.1016/j.jnoncrysol.2022.122086}

\bibitem{Mattos2022}
G.~R. Mattos, C.~D. Bordon, O.~C. Vilela, L.~A. G{\'{o}}mez-Malag{\'{o}}n,
  L.~R. Kassab, {Enhancement of multijunction solar cell efficiency using a
  cover layer of Eu3+, Tb3+ and Eu3+/Tb3+ doped GeO2-PbO-Al2O3 glasses as
  spectral converter of solar radiation}, Optical Materials 132~(August) (2022)
  1--7.
\newblock \href {https://doi.org/10.1016/j.optmat.2022.112833}
  {\path{doi:10.1016/j.optmat.2022.112833}}.

\bibitem{Yang2014}
F.~Yang, C.~Liu, D.~Wei, Y.~Chen, J.~Lu, S.~E. Yang, {Er3+-Yb3+ co-doped
  TeO2-PbF2 oxyhalide tellurite glasses for amorphous silicon solar cells},
  Optical Materials 36~(6) (2014) 1040--1043.
\newblock \href {https://doi.org/10.1016/j.optmat.2014.01.020}
  {\path{doi:10.1016/j.optmat.2014.01.020}}.

\bibitem{Kadam2021}
A.~R. Kadam, S.~J. Dhoble,
  \href{https://doi.org/10.1016/j.jallcom.2021.161138}{{Energy transfer
  mechanism of KAlF4:Dy3+, Eu3+ co-activated down-conversion phosphor as
  spectral converters: An approach towards improving photovoltaic efficiency by
  downshifting layer}}, Journal of Alloys and Compounds 884 (2021) 161138.
\newblock \href {https://doi.org/10.1016/j.jallcom.2021.161138}
  {\path{doi:10.1016/j.jallcom.2021.161138}}.
\newline\urlprefix\url{https://doi.org/10.1016/j.jallcom.2021.161138}

\bibitem{Jia2018}
H.~Jia, Z.~Liu, L.~Liao, Y.~Gu, C.~Ding, J.~Zhao, W.~Zhang, X.~Hu, X.~Feng,
  Z.~Chen, X.~Liu, J.~Qiu, {Upconversion Luminescence from Ln3+(Ho3+,Pr3+)
  Ion-Doped BaCl2 Particles via NIR Light of Sun Excitation}, Journal of
  Physical Chemistry C 122~(17) (2018) 9606--9610.
\newblock \href {https://doi.org/10.1021/acs.jpcc.8b02434}
  {\path{doi:10.1021/acs.jpcc.8b02434}}.

\bibitem{Allsopp2018}
B.~L. Allsopp, G.~Christopoulou, A.~Brookfield, S.~D. Forder, P.~A. Bingham,
  \href{https://www.ingentaconnect.com/content/sgt/ejgst/2018/00000059/00000004/art00005}{{Optical
  and structural properties of d0 ion-doped silicate glasses for photovoltaic
  applications}}, Physics and Chemistry of Glasses: European Journal of Glass
  Science and Technology Part B 59~(4) (2018) 193--202.
\newblock \href {https://doi.org/10.13036/17533562.59.4.003}
  {\path{doi:10.13036/17533562.59.4.003}}.
\newline\urlprefix\url{https://www.ingentaconnect.com/content/sgt/ejgst/2018/00000059/00000004/art00005}

\bibitem{Fujita2018}
K.~Fujita, R.~Watanabe, Y.~Iso, T.~Isobe,
  \href{https://doi.org/10.1016/j.jlumin.2018.02.023}{{Preparation and
  characterization of Y2O3:Bi3+,Yb3+ nanosheets with wavelength conversion from
  near-ultraviolet to near-infrared}}, Journal of Luminescence 198~(January)
  (2018) 243--250.
\newblock \href {https://doi.org/10.1016/j.jlumin.2018.02.023}
  {\path{doi:10.1016/j.jlumin.2018.02.023}}.
\newline\urlprefix\url{https://doi.org/10.1016/j.jlumin.2018.02.023}

\bibitem{Peng2009}
M.~Peng, L.~Wondraczek, {Bismuth-doped oxide glasses as potential solar
  spectral converters and concentrators}, Journal of Materials Chemistry 19~(5)
  (2009) 627--630.
\newblock \href {https://doi.org/10.1039/b812316k}
  {\path{doi:10.1039/b812316k}}.

\bibitem{Ghosh2015}
D.~Ghosh, S.~Balaji, K.~Biswas, K.~Annapurna, {Broad NIR emission near c - Si
  band gap from Bi-doped Ba–Al metaphosphate glasses as promising solar
  spectral converter}, Journal of Materials Science 50~(16) (2015) 5450--5457.
\newblock \href {https://doi.org/10.1007/s10853-015-9090-1}
  {\path{doi:10.1007/s10853-015-9090-1}}.

\bibitem{Svrcek2004}
V.~{\v{S}}vr{\v{c}}ek, A.~Slaoui, J.~C. Muller, {Silicon nanocrystals as light
  converter for solar cells}, Thin Solid Films 451-452 (2004) 384--388.
\newblock \href {https://doi.org/10.1016/j.tsf.2003.10.133}
  {\path{doi:10.1016/j.tsf.2003.10.133}}.

\bibitem{Luxembourg2014}
S.~L. Luxembourg, A.~R. Burgers, R.~Limpens, T.~Gregorkiewicz, A.~Weeber,
  \href{http://dx.doi.org/10.1016/j.egypro.2014.08.115}{{Application of a
  silicon nanocrystal down-shifter to a c-Si solar cell}}, Energy Procedia 55
  (2014) 190--196.
\newblock \href {https://doi.org/10.1016/j.egypro.2014.08.115}
  {\path{doi:10.1016/j.egypro.2014.08.115}}.
\newline\urlprefix\url{http://dx.doi.org/10.1016/j.egypro.2014.08.115}

\bibitem{Yuan2020}
Z.~Yuan, T.~Nakamura, {Spectral tuning of colloidal Si nanocrystal luminescence
  by post-laser irradiation in liquid}, RSC Advances 10~(54) (2020)
  32992--32998.
\newblock \href {https://doi.org/10.1039/d0ra05205a}
  {\path{doi:10.1039/d0ra05205a}}.

\bibitem{Das2021}
B.~Das, S.~M. Hossain, A.~Nandi, D.~Samanta, A.~K. Pramanick, S.~O.
  {Mart{\'{i}}nez Chapa}, M.~Ray, {Spectral conversion by silicon nanocrystal
  dispersed gel glass: Efficiency enhancement of silicon solar cell}, Journal
  of Physics D: Applied Physics 55~(2) (2021).
\newblock \href {https://doi.org/10.1088/1361-6463/ac29e3}
  {\path{doi:10.1088/1361-6463/ac29e3}}.

\bibitem{Sandrini2023}
M.~Sandrini, J.~C. Gemelli, M.~S. Gibin, V.~S. Zanuto, R.~F. Muniz, F.~S.
  de~Vicente, M.~P. Belan{\c{c}}on,
  \href{https://doi.org/10.1016/j.jnoncrysol.2022.122033
  https://linkinghub.elsevier.com/retrieve/pii/S0022309322006275}{{Synthesis
  and properties of Cerium-doped organic/silica xerogels: A potential UV filter
  for photovoltaic panels}}, Journal of Non-Crystalline Solids 600~(July 2022)
  (2023) 122033.
\newblock \href {https://doi.org/10.1016/j.jnoncrysol.2022.122033}
  {\path{doi:10.1016/j.jnoncrysol.2022.122033}}.
\newline\urlprefix\url{https://doi.org/10.1016/j.jnoncrysol.2022.122033
  https://linkinghub.elsevier.com/retrieve/pii/S0022309322006275}

\bibitem{zier2021}
M.~Zier, P.~Stenzel, L.~Kotzur, D.~Stolten, {A review of decarbonization
  options for the glass industry} (jun 2021).
\newblock \href {https://doi.org/10.1016/j.ecmx.2021.100083}
  {\path{doi:10.1016/j.ecmx.2021.100083}}.

\bibitem{griffin2021}
P.~W. Griffin, G.~P. Hammond, R.~C. McKenna, {Industrial energy use and
  decarbonisation in the glass sector: A UK perspective}, Advances in Applied
  Energy 3 (aug 2021).
\newblock \href {https://doi.org/10.1016/j.adapen.2021.100037}
  {\path{doi:10.1016/j.adapen.2021.100037}}.

\bibitem{furszyfer2022}
D.~D. {Furszyfer Del Rio}, B.~K. Sovacool, A.~M. Foley, S.~Griffiths,
  M.~Bazilian, J.~Kim, D.~Rooney, {Decarbonizing the glass industry: A critical
  and systematic review of developments, sociotechnical systems and policy
  options} (mar 2022).
\newblock \href {https://doi.org/10.1016/j.rser.2021.111885}
  {\path{doi:10.1016/j.rser.2021.111885}}.

\bibitem{Flores2019}
L.~F. Flores, K.~Y. Tucto, J.~A. Guerra, J.~A. T{\"{o}}fflinger, E.~S. Serquen,
  A.~Osvet, M.~Batentschuk, A.~Winnacker, R.~Grieseler, R.~Weing{\"{a}}rtner,
  \href{https://doi.org/10.1016/j.optmat.2019.04.003}{{Luminescence properties
  of Yb3+-Tb3+ co-doped amorphous silicon oxycarbide thin films}}, Optical
  Materials 92~(March) (2019) 16--21.
\newblock \href {https://doi.org/10.1016/j.optmat.2019.04.003}
  {\path{doi:10.1016/j.optmat.2019.04.003}}.
\newline\urlprefix\url{https://doi.org/10.1016/j.optmat.2019.04.003}

\bibitem{Bubli2020}
I.~Bubli, S.~Ali, M.~Ali, K.~Hayat, Y.~Iqbal, S.~Zulfiqar, A.~ul~Haq,
  E.~Cattaruzza, \href{https://doi.org/10.1016/j.ceramint.2019.09.193
  https://linkinghub.elsevier.com/retrieve/pii/S0272884219327300}{{Enhancement
  of solar cell efficiency via luminescent downshifting by an optimized
  coverglass}}, Ceramics International 46~(2) (2020) 2110--2115.
\newblock \href {https://doi.org/10.1016/j.ceramint.2019.09.193}
  {\path{doi:10.1016/j.ceramint.2019.09.193}}.
\newline\urlprefix\url{https://doi.org/10.1016/j.ceramint.2019.09.193
  https://linkinghub.elsevier.com/retrieve/pii/S0272884219327300}

\bibitem{Bouajaj2016}
A.~Bouajaj, S.~Belmokhtar, M.~Britel, C.~Armellini, B.~Boulard, F.~Belluomo,
  A.~{Di Stefano}, S.~Polizzi, A.~Lukowiak, M.~Ferrari, F.~Enrichi,
  \href{http://dx.doi.org/10.1016/j.optmat.2015.12.013
  https://linkinghub.elsevier.com/retrieve/pii/S0925346715301543}{{Tb3+/Yb3+
  codoped silica–hafnia glass and glass–ceramic waveguides to improve the
  efficiency of photovoltaic solar cells}}, Optical Materials 52 (2016) 62--68.
\newblock \href {https://doi.org/10.1016/j.optmat.2015.12.013}
  {\path{doi:10.1016/j.optmat.2015.12.013}}.
\newline\urlprefix\url{http://dx.doi.org/10.1016/j.optmat.2015.12.013
  https://linkinghub.elsevier.com/retrieve/pii/S0925346715301543}

\bibitem{Enrichi2021}
F.~Enrichi, E.~Cattaruzza, P.~Riello, G.~C. Righini, A.~Vomiero,
  \href{https://doi.org/10.1016/j.ceramint.2021.03.107}{{Ag-sensitized
  Tb3+/Yb3+ codoped silica-zirconia glasses and glass-ceramics: Systematic and
  detailed investigation of the broadband energy-transfer and downconversion
  processes}}, Ceramics International 47~(13) (2021) 17939--17949.
\newblock \href {https://doi.org/10.1016/j.ceramint.2021.03.107}
  {\path{doi:10.1016/j.ceramint.2021.03.107}}.
\newline\urlprefix\url{https://doi.org/10.1016/j.ceramint.2021.03.107}

\bibitem{Langenhorst2019c}
M.~Langenhorst, D.~Ritzer, F.~Kotz, P.~Risch, S.~Dottermusch, A.~Roslizar,
  R.~Schmager, B.~S. Richards, B.~E. Rapp, U.~W. Paetzold, {Liquid Glass for
  Photovoltaics: Multifunctional Front Cover Glass for Solar Modules}, ACS
  Applied Materials and Interfaces 11~(38) (2019) 35015--35022.
\newblock \href {https://doi.org/10.1021/acsami.9b12896}
  {\path{doi:10.1021/acsami.9b12896}}.

\bibitem{Wang2022}
P.~Wang, X.~Yan, H.~Wang, C.~Luo, C.~Wang,
  \href{https://doi.org/10.1016/j.optmat.2022.112821}{{Study on improving the
  efficiency of crystalline silicon photovoltaic module with down-conversion
  chlorophyll film}}, Optical Materials 132~(May) (2022) 112821.
\newblock \href {https://doi.org/10.1016/j.optmat.2022.112821}
  {\path{doi:10.1016/j.optmat.2022.112821}}.
\newline\urlprefix\url{https://doi.org/10.1016/j.optmat.2022.112821}

\bibitem{Bataille2018}
C.~Bataille, M.~{\AA}hman, K.~Neuhoff, L.~J. Nilsson, M.~Fischedick,
  S.~Lechtenb{\"{o}}hmer, B.~Solano-Rodriquez, A.~Denis-Ryan, S.~Stiebert,
  H.~Waisman, O.~Sartor, S.~Rahbar,
  \href{https://linkinghub.elsevier.com/retrieve/pii/S0959652618307686}{{A
  review of technology and policy deep decarbonization pathway options for
  making energy-intensive industry production consistent with the Paris
  Agreement}}, Journal of Cleaner Production 187 (2018) 960--973.
\newblock \href {https://doi.org/10.1016/j.jclepro.2018.03.107}
  {\path{doi:10.1016/j.jclepro.2018.03.107}}.
\newline\urlprefix\url{https://linkinghub.elsevier.com/retrieve/pii/S0959652618307686}

\bibitem{Belancon2020}
M.~P. Belan{\c{c}}on,
  \href{https://linkinghub.elsevier.com/retrieve/pii/S1755008421000016}{{Brazil
  electricity needs in 2030: Trends and challenges}}, Renewable Energy Focus
  36~(00) (2021) 89--95.
\newblock \href {https://doi.org/10.1016/j.ref.2021.01.001}
  {\path{doi:10.1016/j.ref.2021.01.001}}.
\newline\urlprefix\url{https://linkinghub.elsevier.com/retrieve/pii/S1755008421000016}

\bibitem{Lunardi2018}
M.~M. Lunardi, J.~P. Alvarez-Gaitan, J.~I. Bilbao, R.~Corkish,
  \href{http://www.intechopen.com/books/solar-panels-and-photovoltaic-materials/a-review-of-recycling-processes-for-photovoltaic-modules}{{A
  Review of Recycling Processes for Photovoltaic Modules}}, Solar Panels and
  Photovoltaic Materials (2018).
\newblock \href {https://doi.org/10.5772/intechopen.74390}
  {\path{doi:10.5772/intechopen.74390}}.
\newline\urlprefix\url{http://www.intechopen.com/books/solar-panels-and-photovoltaic-materials/a-review-of-recycling-processes-for-photovoltaic-modules}

\bibitem{Ren2021}
K.~Ren, X.~Tang, M.~H{\"{o}}{\"{o}}k, {Evaluating metal constraints for
  photovoltaics: Perspectives from China's PV development}, Applied Energy
  282~(October 2020) (2021).
\newblock \href {https://doi.org/10.1016/j.apenergy.2020.116148}
  {\path{doi:10.1016/j.apenergy.2020.116148}}.

\bibitem{Dias2016a}
P.~R. Dias, M.~G. Benevit, H.~M. Veit, {Photovoltaic solar panels of
  crystalline silicon: Characterization and separation}, Waste Management and
  Research 34~(3) (2016) 235--245.
\newblock \href {https://doi.org/10.1177/0734242X15622812}
  {\path{doi:10.1177/0734242X15622812}}.

\bibitem{Svrcek2015}
V.~Svrcek, T.~Yamanari, D.~Mariotti, S.~Mitra, T.~Velusamy, K.~Matsubara, {A
  silicon nanocrystal/polymer nanocomposite as a down-conversion layer in
  organic and hybrid solar cells}, Nanoscale 7~(27) (2015) 11566--11574.
\newblock \href {https://doi.org/10.1039/c5nr02703a}
  {\path{doi:10.1039/c5nr02703a}}.

\end{thebibliography}


\end{document}